\documentclass[pra,showpacs,twocolumn,floatfix,aps]{revtex4}
\usepackage{graphicx}
\usepackage{epstopdf}

\usepackage[utf8x]{inputenc}
\usepackage{color}
\usepackage[dvipsnames]{xcolor}
\usepackage{footnote}
\usepackage{subcaption}
\usepackage{physics}

\usepackage{stackengine,graphicx}

\usepackage{dsfont}
\usepackage{amsmath, amssymb}

\begin{document}



\title{Anisotropy-assisted thermodynamic advantage of a local-spin thermal machine}







\author{Chayan Purkait$^{1}$}
\email{2018phz0001@iitrpr.ac.in}
\author{Suman Chand$^2$}
\author{Asoka Biswas$^{1}$}

\affiliation{$^1$Department of Physics, Indian Institute of Technology Ropar, Rupnagar, Punjab 140001, India}
\affiliation{$^2$Dipartimento di Fisica, Universit\`a degli studi di Genova, 16146 Genova, Italy}

\date{\today}

\begin{abstract}

We study quantum Otto thermal machines with a two-spin working system coupled by anisotropic interaction. Depending on the choice of different parameters, the quantum Otto cycle can function as different thermal machines, including a heat engine, refrigerator, accelerator and heater. We aim to investigate how the anisotropy plays a fundamental role in the performance of the quantum Otto engine (QOE) operating in different time scales. We find that while the engine’s efficiency increases with the increase in anisotropy for the quasistatic operation, quantum internal friction and incomplete thermalization degrade the performance in a finite-time cycle. 
Further, we study the QOE with one of the spins ('local' spin) as the working system. We show that 
the efficiency of such an engine can surpass the standard quantum Otto limit, along with maximum power, thanks to the anisotropy. This can be attributed to quantum interference effects. 
We demonstrate that the enhanced performance of a local-spin QOE originates from the same interference effects, as in a measurement-based QOE for their finite-time operation.

\end{abstract}

\maketitle

\section {INTRODUCTION}


Recent advancements in experimental techniques have made it possible to measure and control systems at the level of a single atom and molecule. This accelerated as the size of the quantum devices has been shrinking rapidly.   Consequently, it becomes imperative to understand the thermodynamics of quantum systems. This demand has led us to study thermal machines (heat engines, refrigerators, heaters, accelerators) at the atomic level \cite{vinjanampathy2016CP,binder2018book}. Several studies have been done in this direction \cite{bhattacharjee2021EPJB}, and it has been shown that non-classical features, viz., quantum coherence \cite{scully2011PNAS,brandner2015NJP,mitchison2015NJP,uzdin2015PRX,correa2014SR,gelbwaser2015SR}, quantum correlation and entanglement \cite{dillenschneider2009EPL,zhang2007PRE,altintas2015PRA,hewgill2018PRA,brask2015PRE,brunner2014PRE,park2013PRL}, and non-thermal baths \cite{scully2003Science,rossnagel2014PRL,huang2012PRE,niedenzu2018NC,de2019PRL} can be exploited to enhance the performance of quantum thermal machines (QTMs).

Finite power is required in various practical applications of quantum technologies. Operating a quantum heat engine (QHE) quasistatically leads to null power generation, which may not be useful in practice. Furthermore, the finite-time operation of the QTMs may exploit genuine non-classical properties in their performances \cite{purkait2023PRE,chand2021PRE}. In fact, various QHE models have been studied for finite times. It has been shown that the non-Markovian character of dynamics can speed up the control of a quantum system and improve the power output of a thermal machine \cite{abiuso2019PRA,das2020PRR}. In other studies, it is found that quantum coherence can be harnessed to increase the power of QHEs \cite{scully2011PNAS,rahav2012PRA,gelbwaser2015SR,scully2003Science,uzdin2015PRX} and the efficiency at maximum power (EMP) \cite{dorfman2018PRE}, as well. Furthermore, the role of quantum internal friction on the work extraction and performance of the QHEs has been investigated \cite{ccakmak2017EPJD,turkpencce2019QIP,ccakmak2019PRE,plastina2014PRL,rezek2010Ent,chand2021PRE}.

Coupled spin systems play an important role as a working system for QTMs \cite{thomas2011PRE,ccakmak2016lEPJP,altintas2015PRE,ccakmak2017EPJP,huang2014EPJP,ivanchenko2015PRE,turkpencce2019QIP,das2019Entropy,chand2017PRE}. The coupling strength between the spin can serve as an additional control parameter for the cycle \cite{dodonov2018JPA}. 
Such a system offers a platform to explore several aspects of quantum information theory \cite{loss1998PRA,meier2003PRL}. The anisotropy in the coupling between the spins adds further flexibility. The effect of such anisotropy on entanglement \cite{kamta2002PRL,rigolin2004IJQI,zhou2003PRA,guo2007CTP,hu2006JPA}, teleportation \cite{rojas2017AP,yeo2005JPA,zhou2008EPJD} and the tripartite uncertainty bound \cite{haddadi2022EPJP} has been studied. Recently, the role of anisotropy in quantum batteries has been studied \cite{ghosh2022PRA,ghosh2020PRA,konar2022ArXiv}. It was shown that the maximum power output of this battery can be enhanced by maintaining the anisotropy at low values. Though there are there are only a few studies on the effects of anisotropy on the performance of quantum thermal machines \cite{ccakmak2016EPJP,purkait2023PRE}.


In this work, we will study the performance of the QOE with a two-spin working system coupled by Heisenberg's anisotropic XY interaction. Our investigation focuses on different time limits of the cycle: firstly the quasistatic operation; secondly the nonadiabatic unitary processes; then thirdly the incomplete thermalization in the hot isochoric process. For anisotropy in the interaction between the spin, the Hamiltonian does not commute at two times, which introduces genuine quantum nature in the finite time operation of the cycle \cite{rezek2010Entropy}. We will investigate how anisotropy affects the engine's performance both for the quasistatic and finite-time operation of the engine. We show that the efficiency increases with the increase of the anisotropy for the quasistatic operation. For the finite-time operation, we show that irreversibility increases with the increase of the anisotropy.

Then we will consider a single-spin working system which is a part of the global two-spin system and let's call it a local system. We will study the heat engine (HE) operation with such a local spin under the effect of another spin, let's call it a local spin HE when the two-spin is coupled. Primarily, we aim to investigate how it differs from a single-spin HE which is not under the effect of another spin. Can we get any thermodynamic advantage under such a local scenario? Several studies have been conducted on QHEs and refrigerators that function with a local system \cite{huang2013PRE, thomas2011PRE, altintas2015PRE, huang2014EPJP, das2019Entropy, cherubim2019Entropy,zhao2017QIP,Sonkar2023PRA,el2022JOPB}. These studies primarily focused on studying the quasistatic operation, and also employed the Hamiltonian that commutes at different times. We want to explore how the anisotropic interaction, therefore the non-commuting nature of the Hamiltonian affects the performance of a local spin HE. We show that for the quasistatic operation of the HE, work extraction locally is more powerful than globally, and also the efficiency of a local spin HE can outperform a single spin HE. We also show that in the finite-time operation, the efficiency can be enhanced further than the quasistatic limit, and the maximum power is associated with the enhanced efficiency.


The paper is organized as follows. We present our HE model and implementation of the cycle in \textbf{Sec.~\ref{Cycle implementation}}. In \textbf{Sec.~\ref{HE operation in different time frames}}, we discuss the various limiting cases of time of the HE operation. Further in \textbf{Sec.~\ref{local spin quantum heat engine}}, we explore the HE operation using a local spin working system. In \textbf{Sec.~\ref{discussion}} discusses potential experimental implementations of our HE model. Finally, we conclude our work in \textbf{Sec.~\ref{conclusion}}.

\section{Implementation of the quantum Otto cycle} \label{Cycle implementation}

\subsection{System model}

We consider a system of two-spin coupled by an anisotropic XY interaction (with anisotropy parameter $0 \leq \gamma \leq 1$) of Heisenberg type in a transverse time-dependent magnetic field $B(t)$. The Hamiltonian that describes this system can be written as (in the unit of $\hbar=1$) \cite{purkait2023PRE,suzuki2012book}
\begin{equation}\label{total H}
    \hat{H}(t)=\hat{H}_{0}(t) + \hat{H}_{I},
\end{equation}
where,
$\hat{H}_0=B(t)(\hat{\sigma}_1^z+\hat{\sigma}_2^z)$ represents the free part, $\hat{H}_I=J\left[(1+\gamma) \hat{\sigma}_1^x \hat{\sigma}_2^x+(1-\gamma) \hat{\sigma}_1^y \hat{\sigma}_2^y\right]$ represents the interaction between two-spin with the coupling strength $J$ and $\hat{\sigma}_{i}^{x,y,z}$ are the Pauli spin operators for the $i$th spin ($i\in 1,2$).
When $\gamma = 0$, the above Hamiltonian describes an isotropic XX interaction and the Ising spin Hamiltonian for $\gamma = 1$. Because $[\hat{H}_0, \hat{H}_I]\neq 0$ for $\gamma \neq 0$, it renders $[\hat{H}(t_1), \hat{H}(t_2)]\neq 0$, that leads to a true quantum feature in the operation of the finite-time QHE \cite{rezek2010Entropy,purkait2023PRE}.





The eigenvalues and the corresponding eigenvectors of the total Hamiltonian (\textbf{Eq.~\ref{total H}}) are given by 
\begin{equation}\label{energy levels}
  \begin{array}{l}
  \ket{\psi_{0,3}} = \frac{1}{\sqrt{2}}(\frac{B \mp k}{\sqrt{k^{2}  \mp Bk}}\ket{11} + \frac{\gamma J}{\sqrt{k^{2}  \mp Bk}} \ket{00}), ~E_{0,4} = \mp 2k\\
  \ket{\psi_{1,2}} = \frac{1}{\sqrt{2}}(\mp \ket{10} + \ket{01}), ~~~~~~~~~~~~~~~~~~~~E_{1,2} = \mp 2J,
\end{array}  
\end{equation}
where $k=\sqrt{B^{2}+\gamma^{2} J^2}$.

\subsection{Bath Model}

To describe the dynamics of the system under a heat bath, the Lindblad master equation in the interaction picture can be obtained as \cite{breuer2002book,huang2012PRE,purkait2023PRE}
\begin{equation}\label{master}
\begin{aligned}
\frac{\partial{\hat{\rho}}}{\partial{t}} = 
  & i[\hat{\rho},\hat{H}(t)] +  \sum_{i=1,2}[\Gamma(n_{i}+1) \times \\
  &(\hat{X}_{i} \hat{\rho} \hat{X}_{i}^{+}-\frac{1}{2} \hat{X}_{i}^{+}\hat{X}_{i} \hat{\rho}-\frac{1}{2} \hat{\rho} \hat{X}_{i}^{+}\hat{X}_{i}) \\
   & + \Gamma n_{i}(\hat{X}_{i}^{+} \hat{\rho} \hat{X}_{i}-\frac{1}{2} \hat{X}_{i} \hat{X}_{i}^{+} \hat{\rho}-\frac{1}{2} \hat{\rho} \hat{X}_{i}\hat{X}_{i}^{+})],
\end{aligned}
\end{equation}
where we have considered only one spin of the coupled two-spin system interacting with a heat bath at temperature $T$ to maintain the simplicity of the master equation. The sum over $i$ represents the number of transitions in the system under the heat bath, and the thermal photon number distribution at the transition frequencies in the bath are $n\left(\omega_{i}\right)=\left[\exp \left(\frac{\hbar \omega_{i}}{k T}\right)-1\right]^{-1}$. Here, $\Gamma$ is the coupling constant between the system and the bath. Similarly, we can consider that each spin interacts with the bath.


The jump operators of the system when only the first spin interacts with the heat bath by $\sigma^{x}$ operator are given by\cite{purkait2023PRE,breuer2002book}
\begin{equation}
 \begin{array}{l}
 X_{1,2}=\frac{1}{2}\left(\frac{B + k \mp \gamma J}{\sqrt{k^{2} + Bk}}\ket{\psi_{1,2}}\bra{\psi_{3}} + \frac{B - k \pm \gamma J}{\sqrt{k^{2} - Bk}}\ket{\psi_{0}}\bra{\psi_{2,1}}\right).
 \end{array}
\end{equation}
They signify transitions in energy of the system $\hbar\omega_{1}=2k+2J$ and $\hbar\omega_{2}=2k-2J$ respectively.

\subsection{Quantum Otto Cycle and thermodynamic quantities}\label{cycle}
\begin{figure}[h!]
\includegraphics[width=0.4\textwidth]{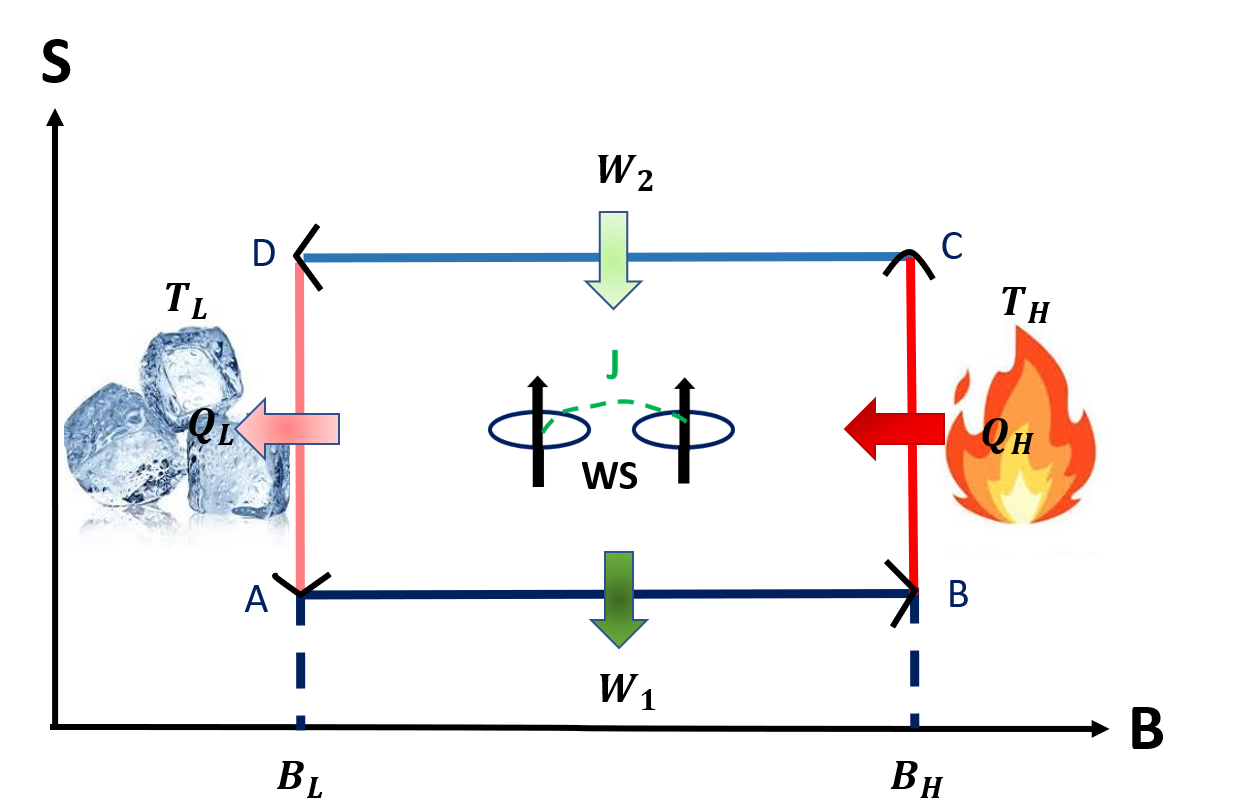}
   \caption{Schematic diagram of the quantum Otto cycle on the entropy ($S$) versus magnetic field ($B$) plane when it functions as a heat engine. In other types of thermal machines, the direction of heat flows and work differ.}
   \label{fig:ottocycle}
\end{figure}


In the following, we will discuss the implementation of the four strokes of the quantum Otto cycle. The schematic diagram of the cycle is shown in \textbf{Fig.~\ref{fig:ottocycle}}.

{\bf Unitary expansion ($A \rightarrow B$)
}: We assume that the cycle begins with the working system in thermal equilibrium with the cold bath at temperature $T_L = 1/\beta_L\left(k_{B}=1\right)$ at point A. The corresponding thermal state of the system is $\hat{\rho}_{A}=e^{-\beta_L \hat{H}_{1}} / Z_{1}$, with $\hat{H}_{1}=\hat{H}(0)$ and $Z_{1}=\operatorname{Tr}(e^{-\beta_{L} \hat{H}^{exp}_{1}})$. In this stroke, the working medium is disconnected from the cold heat bath, and the external magnetic field is changed from $B_{L}$ to $B_{H}$ $(B_{L}<B_{H})$ following the protocol $B(t) = B_{L} + (B_{H} - B_{L}) (t/\tau)$, where $0 \le t \le \tau$ and $\tau$ is the timescale of changing the magnetic field from $B_{L}$ to $B_{H}$ or vice versa. So at point B, the state of the system can be obtained as $\hat{\rho}_{B}=\hat{U}(\tau) \hat{\rho}_{A} \hat{U}^{\dagger}(\tau)$, where $\hat{U}(\tau)=\mathcal{T} e^{-i \int_{0}^{\tau} d t \hat{H}^{exp}(t)}$ is the time evolution operator, $\mathcal{T}$ indicates the time-ordering. 
The amount of work done by the system in this process is given by $W_{1}=\langle E_{B}\rangle - \langle E_{A}\rangle$, where $\langle E_{A}\rangle=\operatorname{Tr}(\hat{\rho}_{A} \hat{H}_{1})$ and  $\langle E_{B}\rangle=\operatorname{Tr}(\hat{\rho}_{B} \hat{H}_{2})$, represent the internal energies of the system at A and B, and $\hat{H}_{2}=\hat{H}(\tau)$ represents the Hamiltonian of the system at B.

{\bf Isochoric heating ($B \rightarrow C$)}: In this stroke, the working medium is connected with a heat bath at temperature $T_{H}$ $(T_{H}>T_{L})$, and the external magnetic field remains fixed at a value $B_{H}$, so the Hamiltonian of the system remains fixed. Therefore, there is no work exchange in this stroke. Also, if the process is carried out for a time $t_{h}$ and the relaxation time of the system is $t_{relax}$, then the case $t_h >> t_{relax}$ represents the system is completely thermalized, 
otherwise, the system is incompletely thermalized in this process. At the end of this process, the state of the system, in the case of complete thermalization, can be represented by  $\hat{\rho}_{C}=e^{-\beta_H \hat{H}_{2}} / Z_{2}$ at temperature $T_H = 1/\beta_H\left(k_{B}=1\right)$, with $\hat{H}_{2}=\hat{H}(\tau)$ and $Z_{2}=\operatorname{Tr}(e^{-\beta_{H} \hat{H}_{2}})$. In the case of incomplete thermalization, the state of the system can be obtained by solving \textbf{Eq.~\ref{master}}. The system absorbs some amount of heat in this process which can be calculated as, $Q_{H} = \langle E_{C}\rangle - \langle E_{B}\rangle$, where $\langle E_{C}\rangle=\operatorname{Tr}(\hat{\rho}_{C} \hat{H}_{2})$ is the internal energy of the system at C. 


{\bf Unitary compression ($C \to D$)}: In this stroke, again the working system is disconnected from the hot heat bath and the external magnetic field is changed from $B_{H}$ to $B_{L}$ following the protocol $B(\tau - t)$, where $0 \le t \le \tau$. In this process, the state of the working system changes to $\hat{\rho}_{D}=\hat{V}(\tau) \hat{\rho}_{C} \hat{V}^{\dagger}(\tau)$, where $\hat{V}(\tau)=\mathcal{T} e^{-i \int_{0}^{\tau} d t \hat{H}^{com}(t)}$ is the time evolution operator with $\hat{H}^{com}(t) = \hat{H}^{exp}(\tau - 1)$. 
The amount of work done on the system in this process can be obtained as $W_{2}=\langle E_{D}\rangle - \langle E_{C}\rangle$, where $\langle E_{D}\rangle=\operatorname{Tr}(\hat{\rho}_{D} \hat{H}_{1})$, represent the internal energy of the system at D.

{\bf Isochoric cooling ($D \to A$)}: In this stroke, the working system is connected with a cold heat bath at temperature $T_{L}$, and the external magnetic field remains fixed at $B_{L}$. 
If the process is carried out for a time $t_{c}$, then the case $t_c >> t_{relax}$ represents that the system reaches thermal equilibrium with the heat bath at the end of this process. The state of the system comes back to the initial state $\rho_A$, 
and the system releases some amount of heat in this process, which can be obtained as, $Q_{L} = \langle E_{A}\rangle - \langle E_{D}\rangle$.

\subsection{Operation of the quantum Otto cycle as different thermal machines}\label{different thermal machines}


\begin{figure*}[t!]
    \centering
    \begin{subfigure}[t]{0.4\textwidth}
        \centering
        \includegraphics[width=\textwidth]{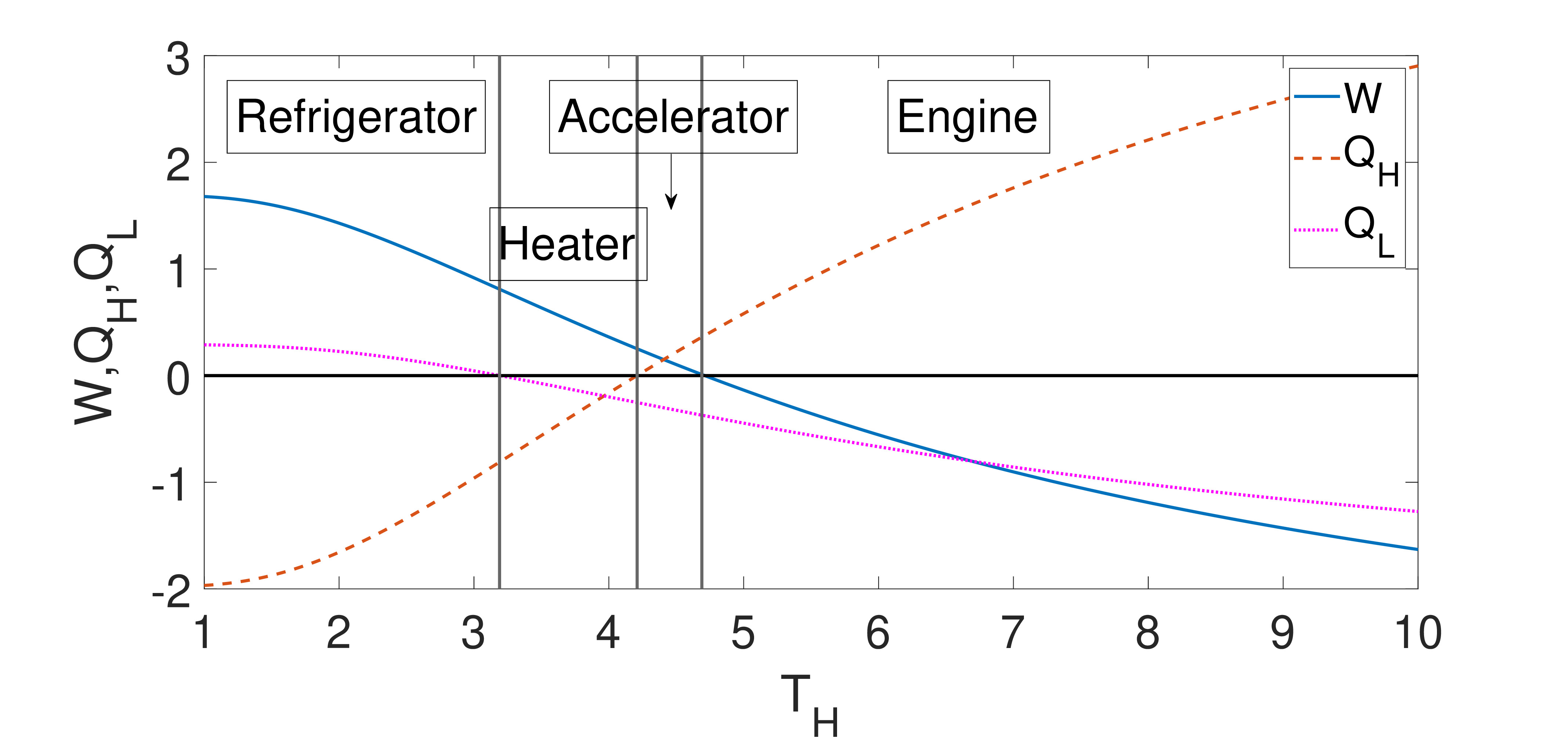}
        \caption{}
    \end{subfigure}%
    ~ 
    \begin{subfigure}[t]{0.4\textwidth}
        \centering
        \includegraphics[width=\textwidth]{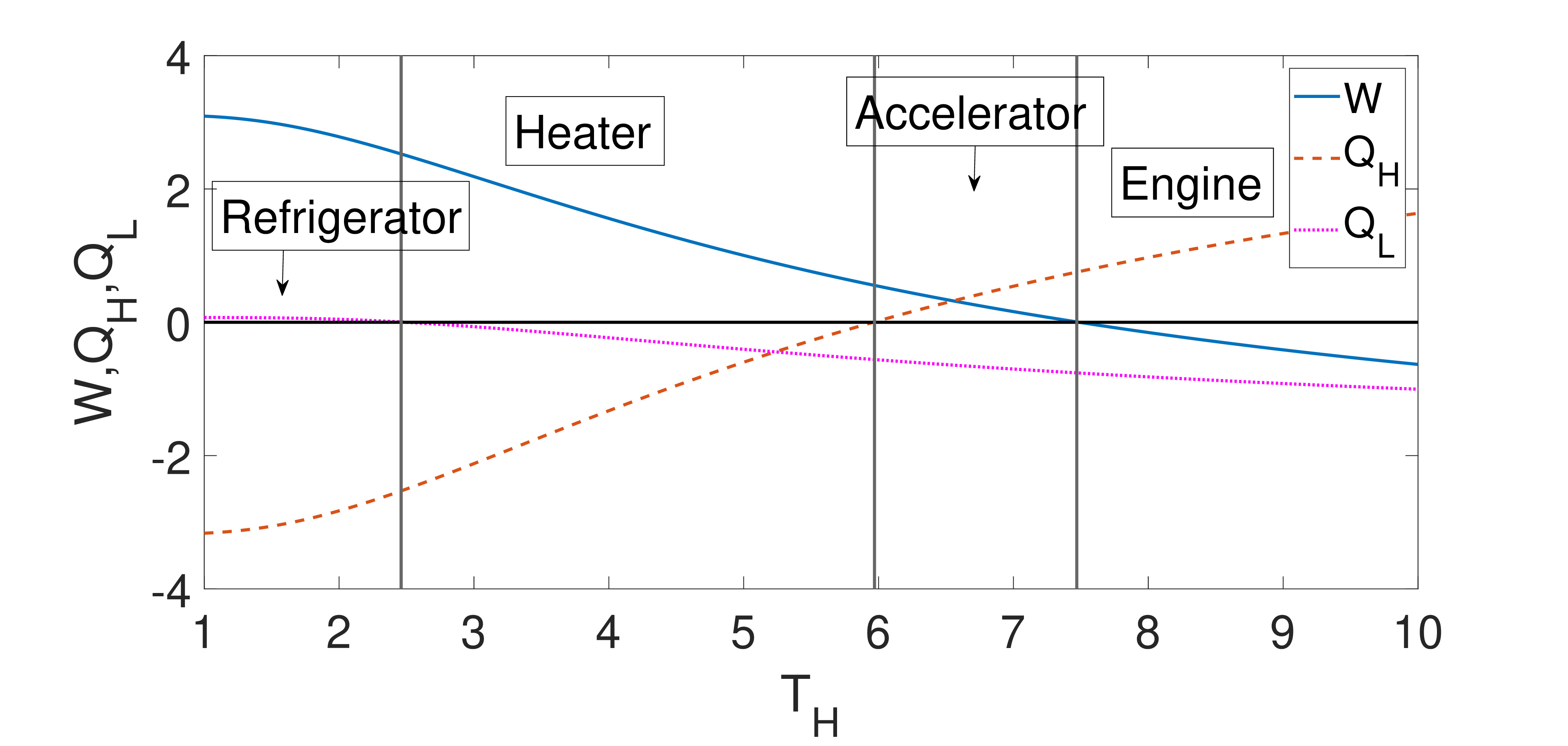}
        \caption{}
    \end{subfigure}
    \caption{Variation of the thermodynamic quantities $W$, $Q_H$ and $Q_L$ as a function of temperature ($T_H$) of the hot bath for different values of the anisotropy parameter (a) $\gamma = 1$, and (b) $\gamma = 0$. The other parameters are $B_L = 1, B_H = 4, J = 1, T_L = 1.$}
    \label{fig:Engine Ref Acc Heater Vs TH}
\end{figure*}

In this part, we will study the parameter regimes of different thermal machines' operation of the cycle \cite{buffoni2019PRL,solfanelli2020PRB}. To do that, we have studied the thermodynamic quantities of the cycle, which are shown in \textbf{Fig.~\ref{fig:Engine Ref Acc Heater Vs TH}}. We observed that with the different choices of the parameters, specifically $T_{H}$ and $\gamma$, the cycle can act in a HE, a refrigerator, an accelerator, or a heater cycle.

The cycle acts as an engine when the system absorbs some amount of heat from the hot bath ($Q_H>0$) and releases a portion of it to the cold bath ($Q_L>0$), and the remaining portion is converted to work ($W<0$) in a complete cycle. It acts as a refrigerator when heat flows in the opposite direction i.e., $Q_L>0$ and $Q_H<0$, with the help of a certain amount of work done on the system ($W>0$). It acts as a thermal accelerator when heat flows in the natural direction, i.e., $Q_H>0$ and $Q_H<0$, as work is done on the system ($W>0$). It operates as a heater when the system releases heat to the hot and cold heat baths, i.e., $Q_H<0$ and $Q_L<0$, with the assistance of work done on the system ($W>0$). From \textbf{Fig.~\ref{fig:Engine Ref Acc Heater Vs TH}}, we can see that the operation regime of different thermal machines varies with the anisotropy parameter $\gamma$.

In our work, we will mainly focus on the HE operation, as this form is more popular than others in terms of the study of thermodynamics. Therefore, the thermodynamic quantities of the HE are as follows. 
Total work in a complete cycle can be obtained as $W = W_1 + W_2 = - (Q_H + Q_L)$. So, the efficiency of the HE is defined as $$\eta = -\frac{ W_{1} + W_{2}}{Q_{H}} = \frac{Q_H + Q_L}{Q_H}$$.

\section{Operation of the heat engine in different time frames}\label{HE operation in different time frames}

In this section, we will focus on the various limiting cases of time over which the HE can be operated.


\subsection{Quasi-static operation}\label{quasi total ststem}

In this part, we consider that two unitary processes (expansion and compression) in the cycle are carried out over a long time such that these processes are adiabatic, i.e., there is no transition between two energy eigenstates. Furthermore, two isochoric processes are carried out for long times, so the system is fully thermalized at the end of these processes. 
Therefore, in these limiting cases of time, the cycle becomes quasi-static.

The analytical expressions of the internal energies [for derivation see \textbf{App.~\ref{Der int en quas}}] of the working system at A, B, C, and D for a quasistatic cycle are given by
\begin{equation}\label{int en quas}
\begin{aligned}
&\langle E_{A,B}\rangle=- 4K_{L,H} \frac{u_1}{Z_1} - 4J \frac{v_1}{Z_1}\;,\\
&\langle E_{C,D}\rangle=-4 K_{H,L} \frac{u_2}{Z_2}-4 J \frac{v_2}{Z_2}\;,\\
\end{aligned}
\end{equation}
where, $K_{L,H} = \sqrt{B_{L,H}^2 + \gamma^2J^2}$, and $Z_1 = 2\cosh(2K_{L}\beta_L) + 2\cosh(2J\beta_L)$ and $Z_2 = 2\cosh(2K_{H}\beta_H) + 2\cosh(2J\beta_H)$ are the partition functions. Also,
$ u_1 = \sinh (2 K_{L} \beta_L)$,   $u_2 = \sinh (2 K_{H} \beta_H)$, $v_1 =  \sinh (2 J \beta_L)$, and $v_2 = \sinh (2 J \beta_H)$.



The thermodynamic quantities of the cycle can be obtained using the \textbf{Eq.~\ref{int en quas}}, and the work in a complete cycle, $W = W_1 + W_2 $ is given by 
\begin{equation}
W = 4(K_L - K_H) \left( \frac{u_1}{Z_1} -  \frac{ u_2}{Z_2} \right).
\end{equation}
Also, heat absorption in the isochoric heating process is given by 
\begin{eqnarray}\label{quas work}
   Q_H = 4K_H \left( - \frac{u_2}{Z_2} -   \frac{ u_1}{Z_1} \right)
+ ~4J \left( -  \frac{ v_2}{Z_2}  + \frac{v_1}{Z_1} \right). 
\end{eqnarray}
Therefore, the expression of efficiency can be obtained as
\begin{equation}\label{quasistatic eff}
   \eta = \frac{-W}{Q_H} =  1 - \frac{K_L (u_1 - u_2)  + J (v_1 - v_2)}{K_H (u_1 - u_2)  + J (v_1 - v_2)},
\end{equation}
From the expression of the efficiency $\eta$ ( Eq.~\ref{quasistatic eff}), we clearly observe that the efficiency depends on the temperatures of both hot and cold baths, $T_{H}$ and $T_{L}$, and also on the value of the magnetic fields $B_{L}$ and $B_{H}$ and the anisotropy parameter $\gamma$. Also, two intermediate energy levels $ |\psi_{1,2} \rangle$ (\textbf{Eq.~\ref{energy levels}}) participate in the engine operation. To compare with the measurement-based QOE in a coupled two-spin system \cite{purkait2023PRE}, the quasistatic efficiency does not depend on the temperate of the cold bath, and also $ |\psi_{1,2} \rangle$ do not participate in the engine operation.

\begin{figure}[h!]
\includegraphics[width=0.45\textwidth]{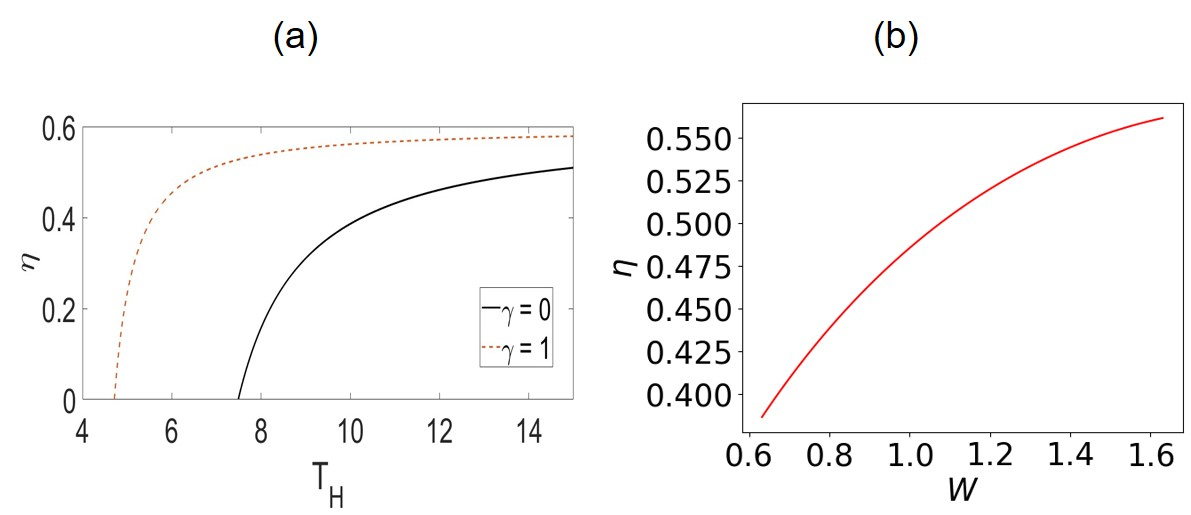}
       \caption {(a) Variation of efficiency ($\eta$) as a function of the temperature ($T_H$) of the hot bath. (b) The parametric plot of the variable anisotropy ($\gamma$) on the work-efficiency plane when $T_H = 10$. $\gamma$ varies from $0$ to $1$, left side of the graph represents $\gamma = 0$ and the right side represents $\gamma = 1$. The other parameters are $B_L = 1, B_H = 4, J = 1, T_L = 1$. The normalization parameter is $J = 1$ throughout this work; therefore, all the quantities are in units of $J$.}
   \label{fig:Engine Effc  Vs TH and Gamma}
\end{figure}

Now, from \textbf{Fig.~\ref{fig:Engine Effc  Vs TH and Gamma}a}, we can observe that the quasistatic efficiency increases gradually and reaches a steady value at higher temperatures of the hot bath $T_{H}$. To operate the engine with higher efficiency, we consider the value $T_H = 10$ in the remaining part of the paper. Further, from the plot of efficiency as a function of work \textbf{Fig.~\ref{fig:Engine Effc  Vs TH and Gamma}b}, we can observe that the work and engine's efficiency both increase with the anisotropy parameter $\gamma$, which is contrary to the measurement-based QOE where quasistatic efficiency decreases with the increase of the anisotropy parameter \cite{purkait2023PRE}. So, we can adjust the parameters of the cycle to achieve a higher efficiency performance of the engine.


\label{fig:Quas Eff Vs TH and Gamma}



\subsection{Unitary time evolution processes are time-dependent}

In this part, we consider that two unitary processes (expansion and compression) in the cycle are time-dependent i.e., in a short time duration it becomes nonadiabatic, and in a large time duration it becomes adiabatic in nature. Also, we consider that the working system is completely thermalized in two isochoric processes i.e., the system is in thermal equilibrium with the baths at the end of the isochoric processes.


\begin{flushleft}
{\bf Thermodynamic quantities in terms of transition probability:}
\end{flushleft}

Under the consideration of the above physical conditions, the expressions of the internal energies [for derivation see \textbf{App.~\ref{Der int en quas}} and \textbf{App.~\ref{der int en finite time}}] of the working system at A, B, C, and D of the cycle are given by
\begin{equation}\label{int ene with tran prob}
\begin{aligned}
\langle E_{A,C}\rangle&=-4 K_{L,H} \frac{u_{1,2}}{Z_{1,2}}-4 J \frac{v_1,v_2}{Z_{1,2}}\;,\\
\langle E_{B,D}\rangle_\tau&=- 4K_{H,L} (1 - 2\xi_\tau) \frac{u_{1,2}}{Z_{1,2}}-4J \frac{v_{1,2}}{Z_{1,2}}\;, \\
\end{aligned}
\end{equation}
where,
$\xi_\tau=|\langle\psi_{0}^{(2)}|\hat{U}(\tau)| \psi_{3}^{(1)}\rangle|^{2} =|\langle\psi_{3}^{(2)}|\hat{U}(\tau)| \psi_{0}^{(1)}\rangle|^{2} =|\langle\psi_{3}^{(1)}|\hat{V}(\tau)| \psi_{0}^{(2)}\rangle|^{2} =|\langle\psi_{0}^{(1)}|\hat{V}(\tau)| \psi_{3}^{(2)}\rangle|^{2}$ represent the transition probability between the energy levels.

Thermodynamic quantities of the HE can be calculated using the \textbf{Eq.~\ref{int ene with tran prob}}. So, work in a complete cycle is given by 
\begin{equation}
    \begin{aligned}
        W_{\tau} = 4K_L \left[ \frac{u_1}{Z_1} -  (1 - 2\xi_\tau) \frac{ u_2}{Z_2} \right]  - 4K_{H} \left [ (1 - 2\xi_\tau) \frac{ u_1}{Z_1}  - \frac{u_2}{Z_2} \right].
    \end{aligned}
\end{equation}
Also, heat absorption in the isochoric heating process is given by 
\begin{equation}
\begin{aligned}
    Q_{\tau} &= 4K_H \left[ - \frac{u_2}{Z_2} -  (1 - 2\xi_\tau) \frac{ u_1}{Z_1} \right] + 4J \left [ -  \frac{ v_2}{Z_2}  + \frac{v_1}{Z_1} \right].
\end{aligned}
\end{equation}
Therefore, the expression of efficiency, $\eta_{\tau} = -W_{\tau}/Q_{\tau}$ is given by
\begin{equation}\label{finite time eff}
\eta_{\tau} = 1 - \frac{K_L \left[ u_1 -  (1 - 2\xi_\tau) u_2 \right]  + J \left [ v_1  - v_2 \right]}{K_H \left[ (1 - 2\xi_\tau) u_1 - u_2 \right]  + J \left [ v_1 -  v_2 \right]}.
\end{equation}


\begin{figure}[h!]
\includegraphics[width=0.46\textwidth]{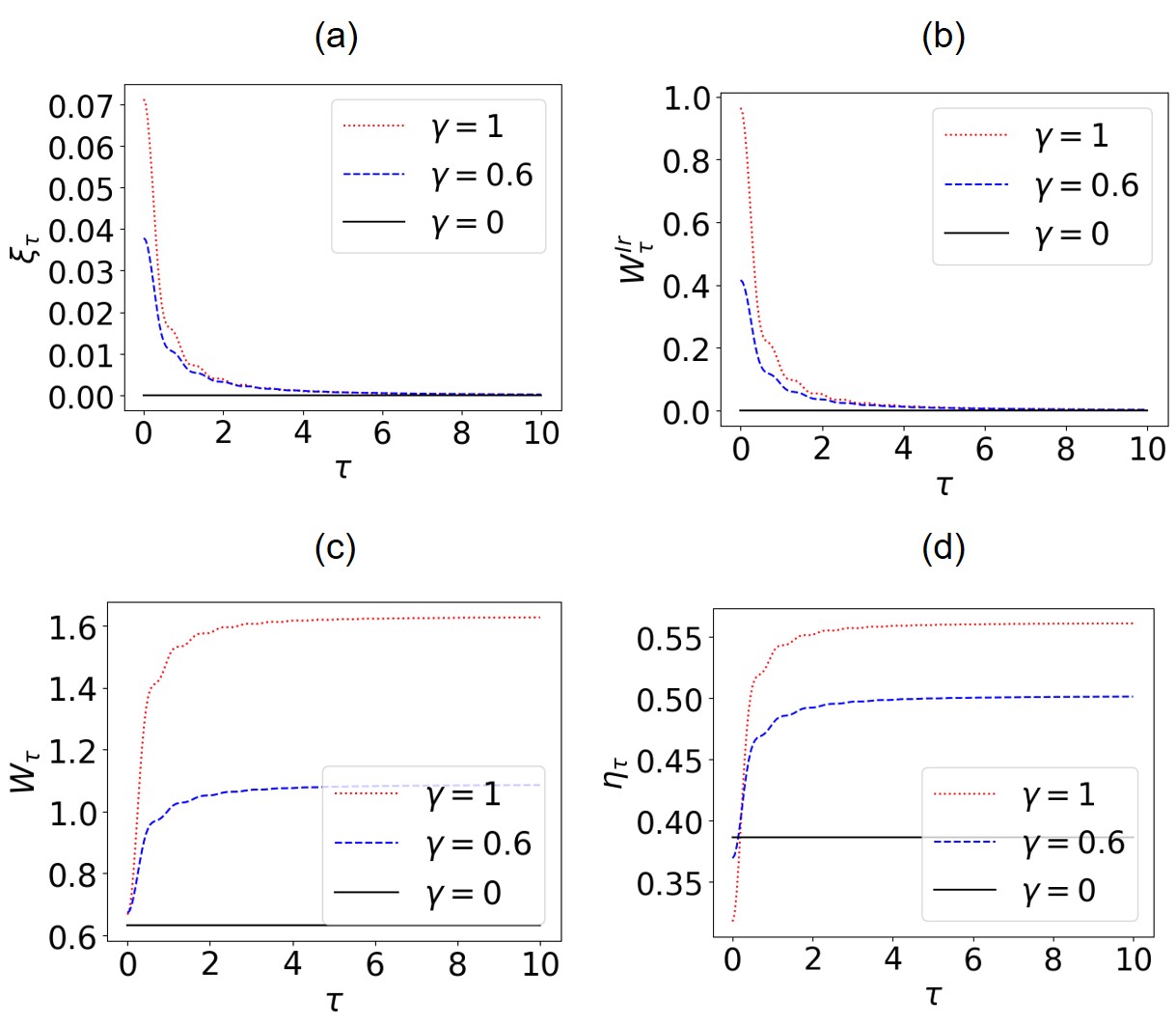}
  \caption{Variation of (a) transition probability ($\xi_\tau$) between two instantaneous energy eigenstates (b) irreversible work $W^{Ir}_{\tau}$ (a) work ($W_{\tau}$) in a complete cycle and (c) efficiency $\eta_{\tau}$  as a function of time of the unitary processes ($\tau$) for different values of anisotropy parameter ($\gamma$). Other parameters are the same with \textbf{Fig.~\ref{fig:Engine Effc  Vs TH and Gamma}}. }
  \label{fig:Xi Wir W Eta Vs unitary time}
\end{figure}


The plot of the transition probability ($\xi_\tau$) as a function of the time of the unitary processes ($\tau$) is shown in \textbf{Fig.~\ref{fig:Xi Wir W Eta Vs unitary time}a}. Now, if we put the value of $\xi_\tau$ in the expression of the finite time efficiency (\textbf{Eq.~\ref{finite time eff}}), we get the plot of efficiency (see \textbf{Fig.~\ref{fig:Xi Wir W Eta Vs unitary time}d}) as a function of $\tau$. In \textbf{Fig.~\ref{fig:Xi Wir W Eta Vs unitary time}c,d}, we have shown how the work in a complete cycle ($W_{\tau}$) and efficiency ($\eta_{\tau}$) varies with $\tau$ for different values of $\gamma$. These plots are produced using QuTip \cite{johansson2012CPC} package. The plots indicate that the engine's work output and efficiency are highly dependent on $\tau$. The work and efficiency both degrade in a very short duration of time and then gradually increase with increasing time and eventually reach the adiabatic (quasistatic) value.





As the Hamiltonian does not commute at different times, the system can not follow the instantaneous energy eigenstates. This induces a nonadiabatic transition between the instantaneous eigenstates of the Hamiltonian when the system is driven by an external control parameter [here $B(t)$], in finite time unitary processes, therefore, unitary processes become nonadiabatic. In this case, work extraction in a complete cycle is reduced. The situation can be seen in this way that an extra amount of work needs to be performed in order to derive the system in finite time, which can be defined by irreversible work
\begin{equation}\label{irr work}
    W^{Ir}_{\tau} = W_{\tau \to \infty} - W_{\tau},
\end{equation}
where $W_{\tau \to \infty}$ is the work which given in  \textbf{Eq.~\ref{quas work}}.
Once the driving process is completed and the system is coupled with the cold bath, the system dumps more amount of heat to the cold bath. 
This degrades the overall performance of the engine in finite-time unitary processes which can be seen in \textbf{Fig.~\ref{fig:Xi Wir W Eta Vs unitary time}c,d}. This is known as quantum internal friction \cite{ccakmak2017EPJD,turkpencce2019QIP,ccakmak2019PRE,plastina2014PRL,rezek2010Ent} and is quantified by $W^{Ir}_{\tau}$. The irreversible work ($W^{Ir}_{\tau}$) represents an irreversibility in the engine performance which is also linked with entropy production in the system in the finite time driving process.


The plot of $W^{Ir}_{\tau}$ with respect to $\tau$ is shown in \textbf{Fig.~\ref{fig:Xi Wir W Eta Vs unitary time}b}. The plot indicates that in the short time limit (nonadiabatic regime), the more the anisotropy ($\gamma$) is more the irreversible work. Therefore, we can say that irreversibility increases with the increase of anisotropy ($\gamma$). For $\gamma = 1$, the system becomes an Ising spin model, which gives rise to maximum irreversibility in finite time operation, and for $\gamma = 0$, the system becomes a Heisenberg XX model which gives rise to reversible operation of the cycle irrespective of the time duration of the unitary processes.

In the adiabatic limit, i.e., $\tau \to \infty$, there is no transition between the instantaneous energy eigenstates. Therefore, in the limit $\tau \to \infty$, we can write $\xi_{\tau} =|\langle\psi_{0}^{(2)}|\hat{U}(\tau)| \psi_{3}^{(1)}\rangle|^{2} \stackrel{\tau \rightarrow \infty}{=} 0 $, which gives rise to $W_\tau = W$, $W_{\tau}^{Ir} = 0$, and $\eta_\tau = \eta$. Therefore, the expression of the quasistatic efficiency (\textbf{Eq.~\ref{quasistatic eff}}) is recovered by putting $\xi_\tau = 0$ in the expression of the finite time efficiency (\textbf{Eq.~\ref{finite time eff}}).

\subsection{Hot isochoric process is time dependent}

In this part, we consider that the hot isochoric process is time-dependent \cite{chand2021PRE,camatiPRA2019}. Therefore, we have different thermalization scenarios of the working system depending on the time limit of this process. In the case $t_{h} >> t_{relax}$, the system is completely thermalized; otherwise, the system is incompletely thermalized. We will investigate how the different time scales, particularly incomplete thermalization, affect the performance of the HE. We also consider that the time for the unitary processes is long enough so that these processes are adiabatic in nature.


With the above-mentioned conditions, the states of the working system at points A and B can be represented by the states as given in \textbf{Eq.~\ref{Der int en quas}} and \textbf{Eq.~\ref{Der int en quas}} respectively. But, to determine the state at point C, we need to solve the master equation (\textbf{Eq.~\ref{master}}), and after that, to determine the state at D, we need to solve the von Neumann equation, which is similar to the situation when there is no dissipative part in \textbf{Eq.~\ref{master}}).

\begin{figure}[h!]
\includegraphics[width=0.46\textwidth]{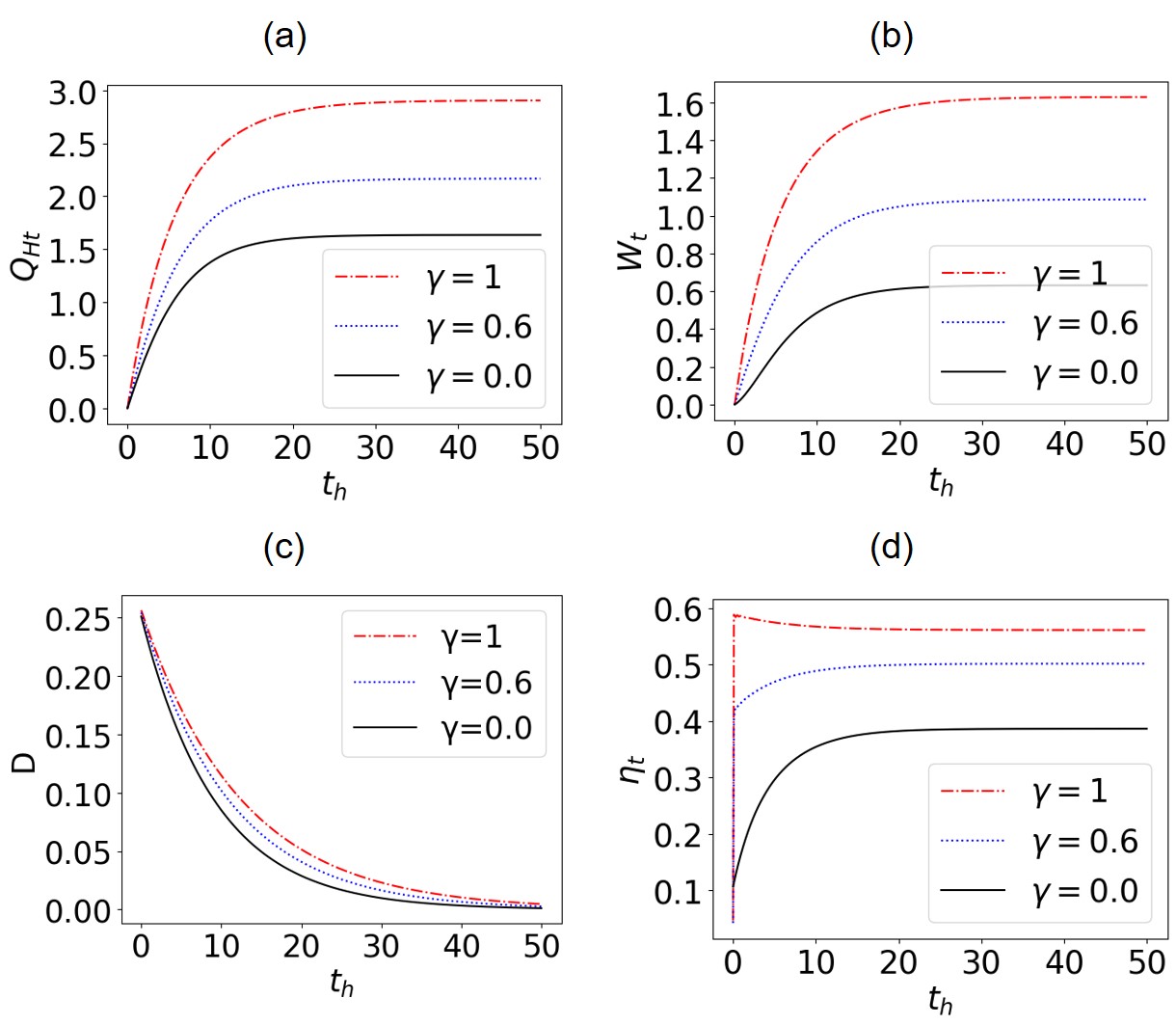}
  \caption{Variation of (a) heat absorption of the system ($Q_{Ht}$) (b) work in a complete cycle (c) trace distance ($D$) between two states, one is for incomplete thermalization and another is the thermal state at C (d) efficiency of the HE as a function time of the isochoric heating process ($t_{h}$) for different values of anisotropy parameter ($\gamma$). For $\gamma = 0$, $D$ is around $75$, whereas, for $\gamma = 1$, $D$ is around $100$, in the unit of $J$, if the accuracy in the trace distance is considered of the order $10^{-5}$. Other parameters are the same with \textbf{Fig.~\ref{fig:Engine Effc  Vs TH and Gamma}} and $\Gamma=0.1$. }
  \label{fig:QH W D Eta Vs th time}
\end{figure}

To understand the thermalization of the working system the trace distance between two states, which is defined as $ D(\rho, \sigma)=\frac{1}{2} \operatorname{Tr}\left|\rho-\sigma\right| $ \cite{camatiPRA2019}, has been studied, where these two states are the reference state represented by \textbf{Eq.~\ref{state C}} and the time-evolved state obtained by solving \textbf{Eq.~\ref{master}}. The plot of the trace distance (D) with respect to the time of the isochoric process ($t_{h}$) is shown in \textbf{Fig.~\ref{fig:QH W D Eta Vs th time}c}. We found that the thermalization time increases with the increase of the anisotropy ($\gamma$) \cite{purkait2023PRE}.




The plots of the heat absorption ($Q_{Ht}$) of the working system from the hot bath and the work done in a complete cycle as a function $t_{h}$ are shown in \textbf{Fig.~\ref{fig:QH W D Eta Vs th time}a,b}. These plots show that the $Q_{Ht}$ increases with the increase of $t_{h}$ and then reaches a steady value when the system is completely thermalized. Also, with the increase in $Q_{Ht}$, the system has more amount of energy to perform work in a complete cycle, therefore the work done increases with $t_{h}$, and reaches a steady at the longer value of $t_{h}$.


The plot of the efficiency ($\eta_{t}$) with respect to $t_{h}$ is shown in \textbf{Fig.~\ref{fig:QH W D Eta Vs th time}d}. For the lower value of $\gamma$, with the time $t_{h}$, work ($W_{t}$) increases slowly than the significant increase of heat absorption ($Q_{Ht}$), which gives rise to a slow increase in
$\eta_{t}$. With increasing $\gamma$, in the very short value of $t_{h}$, $W_{t}$ increases significantly rather than the $Q_{Ht}$, which gives rise to a sudden increase in efficiency. In the large value of $t_h$ both $Q_{Ht}$ and $W_{t}$ become steady, $\eta_{t}$ becomes steady for all values of $\gamma$, which is the quasistatic value of efficiency (see \textbf{Sec.~\ref{quasi total ststem}}).

increases significantly, but in that time period $W_{t}$ also increases but not as much as $Q_{Ht}$ has increased, therefore efficiency decreases in the mid-regime of the time of the isochoric heating.

\section{Heat engine operation of a local system}\label{local spin quantum heat engine}

\begin{figure}[h!]
\includegraphics[width=0.35\textwidth]{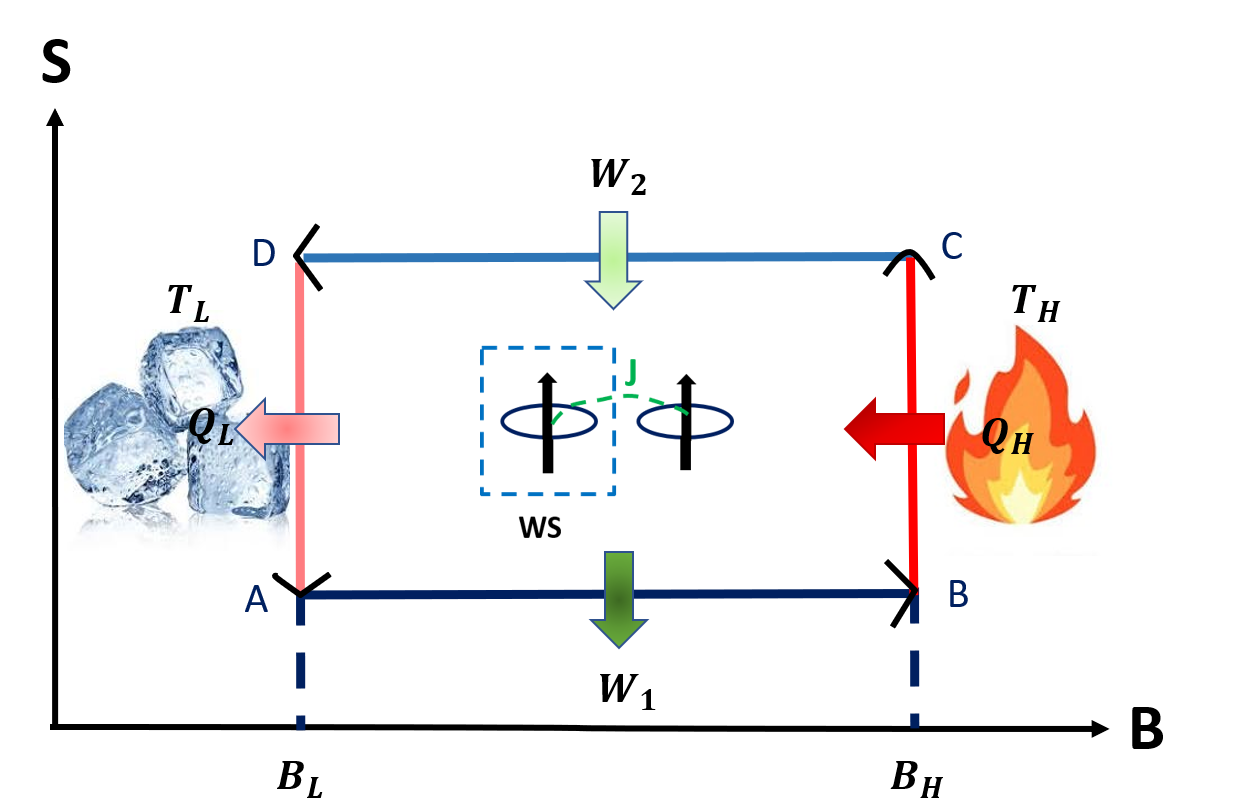}
   \caption{Schematic diagram of the quantum Otto cycle on the entropy ($S$) Vs magnetic field ($B$) plane when it functions as a heat engine.  We consider a single local spin as a working system when the coupled two-spin global system is operated in the Otto cycle. }
   \label{fig:Local Otto cycle}
\end{figure}

In the previous section, we have considered that the coupled two-spin, let's say a global system, operated in the quantum Otto cycle as illustrated in \textbf{Sec.~\ref{cycle}}. In this section, we will consider a single spin which is a part of the global system, let's call it a local system, as a working system. We will study the QOE operation with such a local spin under the effect of another spin. The primary aim is to investigate how the HE operation in a local spin differs from an engine operating with a single-spin working system which is not under the effect of another spin. We want to illustrate the thermodynamic benefits of a local approach in the HE operation.

Quantum heat engines and refrigerators that function with local systems have received significant attention in recent studies \cite{huang2013PRE, thomas2011PRE, altintas2015PRE, huang2014EPJP, das2019Entropy, cherubim2019Entropy,zhao2017QIP,Sonkar2023PRA,el2022JOPB}. These studies primarily focused on analyzing the quasistatic operation of the cycle and also employed the Hamiltonian that commutes at different times. In contrast, our Hamiltonian does not commute at different times (see \textbf{Eq.~\ref{total H}}) which may give rise to some unique characteristics \cite{purkait2023PRE} in the finite time behaviour of the HE operating with a local spin working system. The primary objective is to explore how the non-commuting nature of the Hamiltonian impacts the performance of a local spin HE.




Now to study the thermodynamics of a local spin, we will trace out one spin from the states of the global two-spin system at A, B, C, and D of the cycle (see \textbf{Sec.~\ref{cycle}}), which will give us the states of the local spin. If the states of the global two-spin system are represented by $\rho_j$, where $j \in A, B, C, D$ (see \textbf{App.~\ref{Der int en quas}}, \textbf{App.~\ref{der int en finite time}}), then the reduced density matrices for the first local spin are given by 
$$ \rho_{jL}=\langle 0_{2}|\rho_{j}| 0_{2}\rangle + \langle 1_{2}|\rho_{j}| 1_{2}\rangle, 
$$ 
where subscript 2 represents tracing out the second spin (any spin can be traced out). Therefore, the internal energies of the local spin can be obtained as $\langle E_{j} \rangle_L = \trace (H_{jL} \rho_{jL})$, where $ H_{1L} = B_L \sigma_z$ for $j \in A, D$, and $ H_{2L} = B_H \sigma_z$ for $j \in B, C$ represent the Hamiltonian of the local spin.

Thermodynamic quantities of a local spin can be defined in a similar way as that of the global system (see \textbf{Sec.~\ref{cycle}}). Heat absorption in the isochoric heating process is given by $Q_{HL} = \langle E_{C} \rangle_L - \langle E_{B} \rangle_L$, 
work in the unitary expansion is defined as $W_{1L} = \langle E_{B} \rangle_L - \langle E_{A} \rangle_L$, and that in the unitary compression is defined as $W_{2L} = \langle E_{D} \rangle_L - \langle E_{C} \rangle_L$, so the work in a complete cycle is $W_{L} = W_{1L} + W_{2L}$. 

\subsection{Quasistatic operation of the cycle - }\label{adia local}


Let's consider that the cycle (see \textbf{Sec.~\ref{cycle}}) for the global system is carried out quasistatically, therefore, two unitary processes are adiabatic, and the system is completely thermalized in two isochoric processes. So, the expressions [for derivation see \textbf{App.~\ref{dev int en  local}}] of the internal energies for the local spin are given by 
\begin{equation}\label{int ene local}
\begin{aligned}
&\langle E_{A,D,B,C}\rangle_L=-2 B_{L,L,H,H} (1 - a_{L,L,H,H}^2) \frac{u_{1,2,1,2}}{Z_{1,2,1,2}}\;,
\end{aligned}
\end{equation}
where $a_{L,H} = \frac{B_{L,H} - K_{L,H}}{\sqrt{K_{L,H}^2  - B_{L,H}K_{L,H}}}$.

Thermodynamic quantities of the local spin are given by
\begin{equation}\label{Local W and QH}
  \begin{aligned}
    &W_{L} = 2\left[ B_L(1 - a_L^2) - B_H(1 - a_H^2) \right] \left( \frac{u_1}{Z_1} - \frac{u_2}{Z_2} \right) \\
    &Q_{HL} = 2B_H (1 - a_H^2) \left( \frac{u_1}{Z_1} - \frac{u_2}{Z_2} \right).
\end{aligned}  
\end{equation}










\begin{figure}[h!]
{\includegraphics[width=0.35\textwidth]{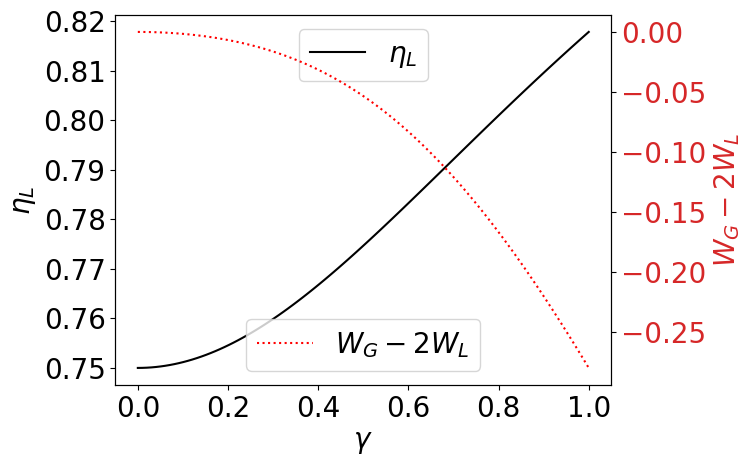}}
  \caption{Variation of the work difference $W_G - 2W_L$ as a function of anisotropy parameter $\gamma$. The figure in the inset represents the variation of efficiency for a local system as a function of the anisotropy parameter ($\gamma$). The efficiency of a single spin system QOE is $0.75$ for the values of $B_L = 1, B_H = 4$. Other parameters are the same with \textbf{Fig.~\ref{fig:Engine Effc  Vs TH and Gamma}}.}
  \label{fig:Local work and eff}
\end{figure}

\begin{flushleft}
{\bf Comparison between global and local work extraction:}
\end{flushleft}
Now to find the potential figure of merit of the local approach, we will compare the local work extraction with the global work extraction for the two-spin system. To do that, we will study the quantity $W_G - 2W_L$, where $W_G$ (\textbf{Eq.~\ref{quas work}}) represents the work for the global two-spin system and $W_L$ (\textbf{Eq.~\ref{Local W and QH}}) represents the work for a local spin and the multiplication factor $2$ comes to consider the contribution from the two local spin. The quantity $W_G - 2W_L$ can be calculated as
\begin{equation}
\begin{aligned}
&W_G - 2W_L = 4 \left[ (K_H - B_H) - (K_L - B_L) + \right. \\
&\left.(B_Ha_H^2 - B_La_L^2) \right]  \times \left( \frac{\sinh2K_L \beta_L}{Z_1} - \frac{\sinh2K_H \beta_H}{Z_2} \right).
\end{aligned}  
\end{equation}


The variation of $W_G - 2W_L$ with respect to $\gamma$ is shown in \textbf{Fig.~\ref{fig:Local work and eff}}. The plot shows that $W_G < 2W_L$ if the two-spin is coupled by anisotropic interaction. For the isotropic interaction, i.e. in the limit of $\gamma \to 0$, $K_H \to B_H$, $K_L \to B_L$, $a_H^2 \to 0$, and also $a_L^2 \to 0$, so $W_G - 2W_L = 0$. The case $\gamma>0$ gives rise to $(K_H - B_H) < (K_L - B_L)$ and also $a_H^2 < a_L^2$, so $W_G - 2W_L < 0$ i.e., the sum of the local work from each local spin surpasses the global work from the global system. Therefore, we can say that extracting work locally is better than globally in the OOE operation with a two-spin system coupled by anisotropic interaction .

\begin{flushleft}
{\bf Comparison between the efficiencies of a local spin and a single QOE:}
\end{flushleft}

The efficiency of the QOE cycle followed by the local spin $\eta_{L} = -\frac{W_{L}}{Q_{HL}} $, is given by 
\begin{equation}\label{quas local eff}
\eta_{Lq} = 1-\frac{B_L\left(1- a_L^2\right)}{B_H\left(1- a_H^2\right)}.
\end{equation}
The expression of the efficiency of the local spin shows that it depends on $\gamma$ through $a_{L,H}$.

If a QOE operates with a single spin working system under the same physical conditions of $B_L$ and $B_H$ (or the same compression ratio $B_L/B_H$), then the expression of the efficiency of a single spin working system QOE is given by \cite{solfanelli2020PRB,kieu2004PRL}
\begin{equation}\label{single spin eff}
    \eta_S = 1 - \frac{B_L}{B_H}.
\end{equation}

We can see that $\gamma \ge 0 $ makes the quantity $(1 - a_L^2)/(1 - a_H^2) \le 1$, which gives rise to $\eta_{Lq}\ge \eta_S$. Therefore, as $\gamma$ increases the quantity $(1 - a_L^2)/(1 - a_H^2)$ becomes more and more less than $1$, which makes $\eta_{L}$ (\textbf{Eq.~\ref{quas local eff}}) is more and more larger than $\eta_S$ (\textbf{Eq.~\ref{single spin eff}}), and for $\gamma = 0$, we get $\eta_{Lq} = \eta_S$. All of these can be seen in the plot of efficiency (\textbf{Fig.~\ref{fig:Local work and eff}}) of the local spin QOE as a function of $\gamma$. The local spin system QOE outperforms the single spin system QOE for $\gamma > 0$. Therefore, we can say that the efficiency of the QOE operating with a local spin working system in conjunction with another spin with an anisotropic interaction between the spin can surpass the standard quantum Otto limit.

\subsection{Finite time operation: unitary processes are time dependent}

In this section, we consider that two unitary processes in the cycle (see \textbf{Sec.~\ref{cycle}}) for the global two-spin system are carried out in a finite time $\tau$ i.e., they are nonadiabatic in nature. But, the thermalization of the working system in the hot isochoric process is complete. The expressions of the internal energies [for derivation see \textbf{App.~\ref{dev int en local fint}}] of the local spin in terms of transition probabilities are given by \begin{equation}\label{int en local fint}
\begin{aligned}
&\langle E_{A,D}\rangle_{L}= -2B_L(1 - 2\delta_{\tau,\tau \to \infty})\frac{u_{1,2}}{Z_{1,2}}\;,\\
&\langle E_{B,C}\rangle_L= -2B_H(1 - 2\lambda_{\tau,\tau \to \infty})\frac{u_{1,2}}{Z_{1,2}}\;,
\end{aligned}
\end{equation}
where
$
\lambda_\tau=|\langle 00|\hat{U}(\tau)| \psi_3^{(1)}\rangle|^2=|\langle 11|\hat{U}(\tau)| \psi_0^{(1)}\rangle|^2,
$
and 
$
\delta_\tau=|\langle 11|\hat{V}(\tau)| \psi_0^{(2)}\rangle|^2=|\langle 00|\Hat{V}(\tau)| \psi_3^{(2)}\rangle|^2
$ 
represent the non-zero overlap between the basis states of a two-spin system and the instantaneous energy eigenstates. In the adiabatic limit i.e. $\tau \to \infty$, $\lambda_\tau$ and $\delta_\tau$ become $\lambda_{\tau \to \infty} = a_H^2/2$ and $\delta_{\tau \to \infty} = a_L^2/2$ respectively, illustrating that finite time average internal energies (see \textbf{Eq.~\ref{int en local fint}}) approach quasistatic average internal energies (see \textbf{Eq.~\ref{int ene local}}).




Thermodynamic quantities of the local spin are given by 
$$
\begin{aligned}
W_{L \tau} = &-2[\frac{u_1}{Z_1} \left[ B_H(1 - 2\lambda_\tau) - B_L (1 - 2\delta_{\tau \to \infty}) \right] \\
&+ \frac{u_2}{Z_2} \left[ B_L(1 - 2\delta_\tau) - B_H (1 - 2\lambda_{\tau \to \infty}) \right] ], \\
Q_{HL\tau} = &-2B_H \left[ \frac{u_2}{Z_2}(1 - 2\lambda_{\tau \to \infty}) - \frac{u_1}{Z_1} (1 - 2\lambda_\tau) \right]. 
\end{aligned}
$$
So, the efficiency, $\eta_{L\tau} = -\frac{W_{L\tau}}{Q_{HL\tau}} $, of the HE cycle experienced by the local spin in finite time is given by 
\begin{equation}\label{eff local}
 \eta_{L\tau}=1-\frac{B_L\left[u_2(1-2 \delta_\tau)-u_1\left(1-2 \delta_{\tau \to\infty}\right)\right]}{B_H\left[u_2\left(1-2 \lambda_{\tau \rightarrow \infty}\right)-u_1(1-2 \lambda_\tau)\right]}.   
\end{equation}
It can be seen that the finite-time local efficiency depends on the temperatures of the heat baths as the coefficients $u_1, u_2$ depend on the temperatures, whereas the quasistatic local efficiency does not depend on the temperatures of the heat baths.



\begin{figure}[h!]
\includegraphics[width=0.35\textwidth]{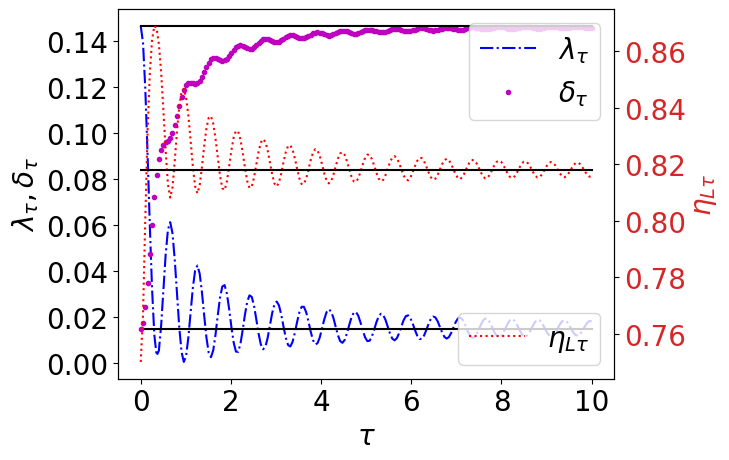}
   \caption{Variation of the transition probability $\lambda_\tau$ and $\delta_\tau$ on the left axis, and efficiency of a local spin on the right axis as a function of time of the unitary processes. The solid line on the top represents the quasistatic value of $\delta_\tau$, at the bottom represents the quasistatic value of $\lambda_\tau$, and in the middle represents the quasistatic value of the local efficiency respectively. Other parameters $\gamma = 1$, remaining are the same with  \textbf{Fig.~\ref{fig:Engine Effc  Vs TH and Gamma}}.}
   \label{fig:prob Eff Vs unitary process time}
\end{figure}

Plots of the transition probabilities ($\lambda, \delta$) with respect to $\tau$ are shown in \textbf{Fig.~\ref{fig:prob Eff Vs unitary process time}}. If we put the value of $\lambda_\tau$ and $\delta_\tau$ in the expression of efficiency (\textbf{Eq.~\ref{eff local}}), we get the plot of efficiency with respect to $\tau$ which is shown in \textbf{Fig.~\ref{fig:prob Eff Vs unitary process time}}. This plot shows that there is an oscillatory dependence of efficiency on $\tau$ for $\gamma \neq 0$. Depending on the exact value of $\tau$ in the short time duration, a local spin system QHE can either underperform or outperform the counterpart which operates in the adiabatic limit. Thus, by adjusting the time of the unitary processes, the efficiency of a local spin system QOE can be enhanced beyond its quasistatic limit. In a long time duration i.e. in the adiabatic limit $(\tau \to \infty)$, efficiency gradually approaches the adiabatic (quasistatic) value (see \textbf{Sec.~\ref{adia local}}). In that case, the local spin system efficiency which is represented by \textbf{Eq.~\ref{eff local}} will be reduced to \textbf{Eq.~\ref{quas local eff}}.

In the sudden quench limit i.e. $\tau \to 0$, the external magnetic field is changed $B_L$ to $B_H$ or vice-versa suddenly, in this case, both the $\delta_\tau$ and $\lambda_\tau$ attain their sudden value which can be obtained as 
$
\lambda_{\tau \to 0}=|\langle 00| \psi_3^{(1)}\rangle|^2=|\langle 11|\psi_3^{(1)}\rangle|^2,
$
and 
$
\delta_{\tau \to 0}=|\langle 11| \psi_3^{(2)}\rangle|^2=|\langle 00| \psi_3^{(2)}\rangle|^2,
$ 
as in this case $\hat{U}(\tau), \hat{V}(\tau) \rightarrow \mathds{1}$. The engine's performance degraded in this case (see \textbf{Fig.~\ref{fig:prob Eff Vs unitary process time}}). Also, in the adiabatic limit i.e. $\tau \to \infty$, both $\lambda_\tau$ and $\delta_\tau$ reach their adiabatic value $\lambda_{\tau \to \infty}$ and $\delta_{\tau \to \infty}$. In between these two limiting cases of time, there is an oscillation in $\delta_\tau, \lambda_\tau$ with respect to $\tau$. The oscillation in the efficiency is mainly because of the oscillation in the transition probabilities $\delta_\tau, \lambda_\tau$ in finite times of the unitary processes, which can be attributed to the interference-like phenomena that happen between two probability amplitudes, which can be seen if we rewrite the $\lambda_{\tau}, \delta_{\tau}$ in the form given in \textbf{Eq..~\ref{delta lambda}}.
\begin{equation}\label{delta lambda}
\begin{aligned}
    \lambda_\tau, \delta_\tau = &\left|\frac{\sqrt{2} a_{H,L}}{a_{H,L} d_{H,L}-b_{H,L}c_{H,L}} \langle\psi_3^{(2)}|\hat{U}(\tau),\hat{V}(\tau)| \psi_3^{(1)}\rangle\right. \\ 
    &-\left.\frac{\sqrt{2} c_{H,L}}{a_{H,L} d_{H,L}-b_{H,L} c_{H,L}}\langle\psi_0^{(2)}|\Hat{U}(\tau),\Hat{V}(\tau)| \psi_3^{(1)}\rangle\right|^2,
\end{aligned}
\end{equation}
where $b_{H,L} = \frac{\gamma J}{\sqrt{k_{H,L}^{2}  - B_{H,L}k_{H,L}}}$, $c_{H,L} = \frac{B_{H,L} + k_{H,L}}{\sqrt{k_{H,L}^{2}  + B_{H,L}k_{H,L}}}$, $d_{H,L} = \frac{\gamma J}{\sqrt{k_{H,L}^{2}  + B_{H,L}k_{H,L}}}$. Although the oscillation in $\delta_\tau$ is less prominent here compare to $\lambda_\tau$ in the parameter value we are using for the engine operation (see \textbf{Fig.~\ref{fig:prob Eff Vs unitary process time}}), in other regions of the parameter, particularly $B_H$, the oscillation in $\delta_\tau$ can be found significant. From \textbf{Fig.~\ref{fig:prob Eff Vs unitary process time}} we can see that when $\lambda_\tau$ goes below the $\lambda_{\tau \to \infty}$, the finite time HE outperforms the counterparts operating in the adiabatic limit ($\tau \to \infty$). Also, it can be shown that for $\gamma = 0$ efficiency does not change with $\tau$, which is because there is no interference-like effect in this case \cite{purkait2023PRE}.


\begin{figure}[h!]
  \includegraphics[width=0.35\textwidth]{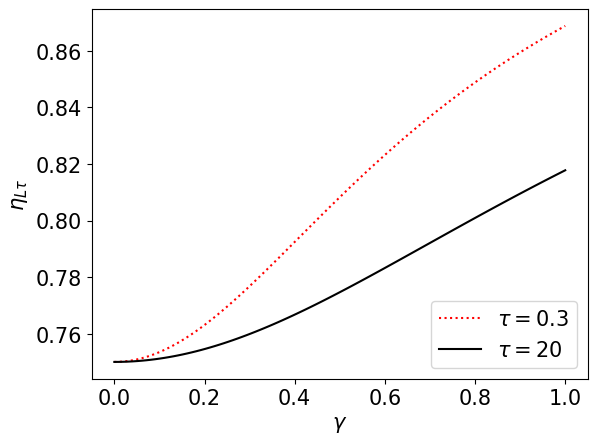}
   \caption{Variation of efficiency of the local spin heat engine as a function of anisotropy parameter ($\gamma$) for different values of the unitary process time ($\tau$). $\tau = 20$ represents the adiabatic and $\tau = 0.3$ represents the non-adiabatic cases of the unitary time evolution. The other parameters are the same with \textbf{Fig.~\ref{fig:Engine Effc  Vs TH and Gamma}}. }
   \label{fig:Eff Vs ansiotropy local}
\end{figure}

The plot of the efficiency of the local spin HE with respect to the anisotropy parameter $\gamma$ is shown in \textbf{Fig.~\ref{fig:Eff Vs ansiotropy local}}. It shows that the outperformance increases with the increase of anisotropy ($\gamma$) for the finite time operation of the engine, which is similar to the measurement-based QOE \cite{purkait2023PRE}.

\begin{flushleft}
{\bf Similarity with a measurement-based QOE: }
\end{flushleft}
It will be worth mentioning if we are able to construct a QHE model with a transition probability between the energy eigenstates and bare basis states of the working system, then we may see an oscillation in the transition probability in finite times. This oscillation allows us to improve the performance of QHEs in finite times than the quasistatic limit. Also. this will be independent of the type of QHE model. In a recent study, it has been shown that the performance of a measurement-based QOE can be enhanced in a finite time using this type of transition probability \cite{purkait2023PRE}. The prescribed type of transition probability is derived from the non-selective measurement protocol. But here we obtain this from a local engine behaviour perspective. Therefore, we can say that the QOE with a local working system can function like a measurement-based engine for the finite-time operation of both of them.

\begin{figure}[h!]
  \includegraphics[width=0.40\textwidth]{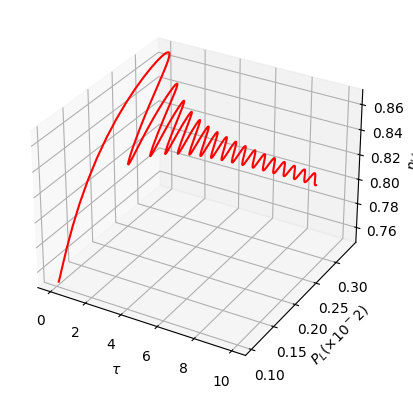}
   \caption{Variation of efficiency ($\eta_{L\tau}$ on the z-axis) of the local spin heat engine as a function of power and the time of the unitary processes ($\tau$) for $\gamma = 1$. Parameters for the isochoric processes are $t_h = 100, t_c = 220$, $\Gamma = 0.1$. Other parameters are the same with \textbf{Fig.~\ref{fig:Engine Effc  Vs TH and Gamma}}.}
   \label{fig:Local Eff Vs local power}
\end{figure}

\begin{flushleft}
{\bf Power Analysis: }
\end{flushleft}
As we are studying the finite-time performance of the engine, it is imperative to explore the power of the engine and its relation to efficiency in this type of local spin HE. The power of the local spin HE can be defined as
\begin{equation}
    P_L = \frac{|W_L|}{t_h + t_c + 2\tau},
\end{equation}
where it is assumed that two isochoric processes are carried out over a long time, but not infinite time so that the states of the working system reach very close to the reference thermal states in two isochoric processes. The 3D plot of  efficiency as a function of power and the time of the unitary processes is shown in \textbf{Fig.~\ref{fig:Local Eff Vs local power}}. From this plot, it can be seen that we can have improved efficiency even at maximum power. 


\section{Discussion}\label{discussion}


A similar type of analysis can be done for the refrigerator operation of the cycle. In contrast to the HE operation, it can be shown that the COP of the refrigerator degrades as the anisotropy ($\gamma$) increases for the quasistatic operation of the cycle. The COP also declines when the refrigerator is operated for a finite time, which is similar to an engine.

Also, using the local analysis as that of the HE mentioned above, we can show that the COP of a local spin refrigerator can be enhanced in finite-time unitary processes, which is similar to the local spin HE operation.


Heisenberg's anisotropic XY interaction between two-spin can be constructed using state-of-the-art technologies \cite{georgescu2014RMP}, particularly in NMR systems or trapped ion systems. In a typical trapped ion system, the coupling constant $J$ can range from a few hundred Hz to one kHz \cite{hess2017PTRSA,grass2014EPJQT}. Also, the external magnetic field can be of the order of a few kHz\cite{grass2014EPJQT,monroe2021RMP,monroe2015book}. Therefore, depending on the value of $J$, the time for the unitary processes $\tau$ can range from $2 \mathrm{\text\textmu s}$ to a few $ \text{ms}$. Also, the working system needs to be cooled at $T_L = 50 ~\text{nK}$ and $T_H = 500~ \text{nK}$.

\section{Conclusions}\label{conclusion}

We have studied the quantum Otto cycle with a two-spin working system coupled by anisotropic interaction. The cycle can be operated in different thermal machine cycles, including a heat engine, refrigerator, accelerator and heater depending on different temperatures of the hot bath, for a fixed value of the coupling constant and the cold bath temperature. Among all thermal machines, the quantum Otto engine (QOE) is studied in different time frames. The role of anisotropy on engine performance has been investigated. We found that the engine's efficiency increases with the increase of the anisotropy parameter ($\gamma$) for the quasistatic operation of the cycle.
But, efficiency decreases for finite-time engine operation due to quantum internal friction. 
We found that the decrease in efficiency increases with the increase of $\gamma$, which signifies irreversibility in engine operation which increases with the increase of $\gamma$. In the isochoric heating process, the case of incomplete thermalization of the working system on the thermodynamic quantities is also discussed. We observed that heat absorption and work in a complete cycle both increase with the increase in the time of the process and reach a steady value after a long time.

Further, we studied the QOE performance with a local spin working system, which is obtained by tracing out one spin from the global two-spin system. We found that the combined local work extraction from all the spin is larger than the global work extraction in the two-spin system and the difference between these two types of work extraction increases with $\gamma$. Also, for anisotropic interaction between two-spin ($\gamma > 0$), a local spin QOE outperforms, in terms of efficiency, a single spin QOE when both function quasistatically with the same cycle parameters. We found that the efficiency of the local spin heat engine oscillates for the finite time unitary processes of the global two-spin system. Therefore, a local spin OQE can outperform the same operating in a long time limit and this outperformance in efficiency is also associated with the maximum power output by the engine. We have shown that the oscillation in efficiency of the local spin QOE comes due to the same origin of an interference-like effect between two probability amplitudes as that of a non-selective measurement-based QOE.

\section*{ACKNOWLEDGMENTS} S. C. would like to acknowledge the funding through the NextGenerationEu Curiosity Driven Project “Understanding even-odd criticality” and the European Union-NextGenerationEU through the “Quantum Busses for Coherent Energy Transfer” (QUBERT) project, in the framework of the Curiosity Driven 2021 initiative of the University of Genova.

\appendix





\appendix

\section{Derivati
on of internal energies for the quasistatic case for the global two-spin system}\label{Der int en quas}


\begin{flushleft}
{\bf At A: }
\end{flushleft}
The Hamiltonian at point A of the cycle can be expressed as
$
H_{A} = H_{1} = \sum_{i=0}^3 E_{i}^{(1)}|\psi_{i}^{(1)}\rangle\langle\psi_{i}^{(1)}|
$
where $\{|\psi_{i}^{(1)}\rangle\}$ are the eigenstates of the Hamiltonian $H_{1}$. As we consider that the system at A is in thermal equilibrium with the heat bath, the thermal density matrix is given by $\rho_{A} =  \frac{e^{-\beta H_{1}}}{Z_1} = \sum_{i = 0}^3 P_{i}|\psi_{i}^{(1)}\rangle\langle\psi_{i}^{(1)}|$, 
where ${P_{i} = e^{-\beta E_{i}^{(1)}}/Z_1}$ is the thermal occupation probability of the $i$th eigenstate. So, the average internal energy at point A is given by $\langle E_{A}\rangle = \Tr(H_{1}\rho_{A}) =\sum_{i = 0}^3 P_{i}E_{i}^{(1)} = -4 K_{L} \frac{u_1}{Z_1}-2 J \frac{v_1}{Z_1}$.



\begin{flushleft}
{\bf At B: }
\end{flushleft}
The Hamiltonian at the point B of the cycle can be expressed as 
$
H_{B} = H_{2} = \sum_{i=0}^3 E_{i}^{(2)}|\psi_{i}^{(2)}\rangle\langle\psi_{i}^{(2)}|,
$
where $\{|\psi_{i}^{(2)}\rangle\}$ are the eigenstates of the Hamiltonian $H_{2}$. We consider that the unitary process AB is carried out adiabatically i.e. the system follows the instantaneous eigenstates, so the state of the system at B can be written as $\rho_{B} = \sum_{n} p_{n}^{\mathrm{L}}|\psi^{(2)}_{n}\rangle\langle \psi^{(2)}_{n}|$.

The average internal energy at the point B, $\langle E_{B}\rangle = \Tr(H_{2}\rho_{B})$ is given by 
$$
\begin{aligned}
\langle E_{B}\rangle & = P_{0}E_{0}^{(2)}~+~P_{3}E_{0}^{(2)}~+~ P_{1}E_{1}^{(2)}~+~P_{2}E_{2}^{(2)}~+~\\
&~~~~~~~~~~~~~~~~~~~~~~~~~~P_{0}E_{3}^{(2)}~+~P_{3}E_{3}^{(2)}\\
& = -4 K_{H} \frac{u_1}{Z_1}-4J \frac{\sinh 2J \beta_L}{Z_1}\;,
\end{aligned}
$$


\begin{flushleft}
{\bf At C: }
\end{flushleft}
The thermal density matrix at C is given by
\begin{equation}\label{state C}
    \rho_{C} =  \frac{e^{-\beta H_{2}}}{Z_2} = \sum_{i = 0}^3 P_{i}|\psi_{i}^{(2)}\rangle\langle\psi_{i}^{(2)}|, 
\end{equation}
where ${P_{i} = e^{-\beta E_{i}^{(2)}}/Z_2}$ is the thermal occupation probability of the $i^th$ eigenstate. Similarly to point A, we can derive the expression of average energy at C which is given by $\langle E_{C}\rangle= \Tr(H_{2}\rho_{c}) =-4 K_{H} \frac{u_2}{Z_2}-4 J \frac{v_2}{Z_2}$.

\begin{flushleft}
{\bf At D: }
\end{flushleft}
Similarly to the unitary process AB, we consider that the unitary process CD is also carried out adiabatically. Therefore, the density matrix at the point D can be written as $\rho_{D} = \sum_{n} p_{n}^{\mathrm{H}}|\psi^{(1)}_{n}\rangle\langle \psi^{(1)}_{n}|$.

Similarly to point B, we can derive the average internal energy at point D which is given by
$$
\langle E_{D}\rangle = \Tr(H_{1}\rho_{D}) =- 4K_{L} \frac{ u_2}{Z_2}-4J \frac{\sinh 2J \beta_H}{Z_2}\;,
$$

\section{Derivation of internal energies of the global two-spin system for finite-time unitary processes }\label{der int en finite time}

\begin{flushleft}
{\bf At B: }
\end{flushleft}
The density matrix at point B after the unitary process AB can be obtained as 
$
\rho_{B\tau} = \hat{U}(\tau)\rho_{A}\hat{U}^{\dag}(\tau) = \sum_{i = 0}^3P_{i}\hat{U}(\tau)|\psi_{i}^{(1)}\rangle\langle\psi_{i}^{(1)}|\hat{U}^{\dag}(\tau)\;.
$

The average internal energy at the point B, $\langle E_{B}\rangle = \Tr(H_{2}\rho_{B\tau})$ is given by
\begin{equation}
\begin{aligned}
&\langle E_{B}\rangle_\tau = P_{0}E_{0}^{(2)}(1 - \xi_\tau)~+~P_{3}E_{0}^{(2)}\xi_\tau~+~ P_{1}E_{1}^{(2)}~+~\\
&~~~~~~~~~~~~~~~~~~~~~~~~P_{2}E_{2}^{(2)}~+~P_{0}E_{3}^{(2)}\xi_\tau~+~P_{3}E_{3}^{(2)}(1-\xi_\tau)\\
& = -4 K_{H} (1 - 2\xi_\tau) \frac{u_1}{Z_1}-4J \frac{v_1 \beta}{Z_1},
\end{aligned}
\end{equation}
where we have used the microreversibility condition $|\langle\psi_{0}^{(2)}| \hat{U}(\tau)|\psi_{3}^{(1)}\rangle |^{2} = |\langle\psi_{3}^{(2)}| \hat{U}(\tau)|\psi_{0}^{(1)}\rangle |^{2} = \xi_\tau$ (for proof see \textbf{App.~\ref{micro}}) and $ |\langle\psi_{0}^{(2)}| \hat{U}(\tau)|\psi_{0}^{(1)}\rangle |^{2} = |\langle\psi_{3}^{(2)}| \hat{U}(\tau)|\psi_{3}^{(1)}\rangle |^{2} = 1 - \xi_\tau$. 
In unitary stages for a short time interval $\tau$, nonadiabatic transitions occur between energy eigenstates that are coupled \cite{cherubim2022PRE}. In the present case, such transitions will be induced between the levels $|\psi_{0}\rangle$ and $|\psi_{3}\rangle$. So, the terms like $\langle\psi_{0}^{(2)}| \hat{U}(\tau)|\psi_{1}^{(1)}\rangle$, $\langle\psi_{0}^{(2)}| \hat{U}(\tau)|\psi_{2}^{(1)}\rangle$, $\langle\psi_{3}^{(2)}| \hat{U}(\tau)|\psi_{1}^{(1)}\rangle$ etc. become zero. More details of the proof can be found in \cite{purkait2023PRE}.


\begin{flushleft}
{\bf At D: }
\end{flushleft}

The density matrix at point D after the unitary process CD is given by
$\rho_{D\tau} = \hat{V}(\tau)\rho_{C}\hat{V}^{\dag}(\tau)$. Similarly to point B, we can derive the average internal energy at point D which is given by
\begin{equation}
\langle E_{D}\rangle_\tau = \Tr(H_{1}\rho_{D\tau}) =- 4K_{L} (1 - 2\xi_\tau) \frac{u_2}{Z_2}-4J \frac{v_2}{Z_2},
\end{equation}
where we have used the microreversibility condition $|\langle\psi_{0}^{(2)}| \hat{V}(\tau)|\psi_{3}^{(1)}\rangle |^{2} = |\langle\psi_{3}^{(2)}| \hat{V}(\tau)|\psi_{0}^{(1)}\rangle |^{2} = \xi_\tau$ (for proof see \textbf{App.~\ref{micro}}) and $ |\langle\psi_{0}^{(2)}| \hat{V}(\tau)|\psi_{0}^{(1)}\rangle |^{2} = |\langle\psi_{3}^{(2)}| \hat{V}(\tau)|\psi_{3}^{(1)}\rangle |^{2} = 1 - \xi_\tau$.

\section{Equivalence of the time evolution operators in the unitary expansion and compression processes}\label{equi}

By utilizing the definitions (see \textbf{Sec.~\ref{cycle}}) of
the unitary time evolution operators in the expansion and compression stages, one can obtain the equivalence between them \cite{camatiPRA2019,cherubim2022PRE}. 
$$
\begin{aligned}
\hat{U}(\tau) & =\mathcal{T} \exp \left[-i \int_0^\tau H^{exp}(t) d t\right] \\
& =\mathcal{T} \exp \left[-i \int_0^{-\tau} H^{exp}(-t) d(-t)\right] \\
& =\mathcal{T} \exp \left[-i \int_\tau^{0} H^{exp}(\tau-t') d (\tau - t')\right] \\
& =\mathcal{T} \exp \left[-i \int_0^{\tau} H^{exp}(\tau-t) d t\right] \\
& =\mathcal{T} \exp \left[-i \int_0^\tau H^{com}(t) d t\right] \\
& =\hat{V}(\tau)
\end{aligned}
$$


\section{Proof of the microreversibility conditions for the total two-spin system:}\label{micro}
Using the completeness relation $\sum_{i=0}^3|\psi_i^{(1)}\rangle \langle \psi_i^{(1)}| = \mathbb{I}$,
and the conservation of probability $|\langle\psi_{0}^{(2)}| \hat{U}(\tau)|\psi_{3}^{(1)}\rangle |^{2} + |\langle\psi_{3}^{(2)}| \hat{U}(\tau)|\psi_{3}^{(1)}\rangle |^{2} = 1$, we can proof the relation $|\langle\psi_{3}^{(2)}| \hat{U}(\tau)|\psi_{0}^{(1)}\rangle |^{2} =  |\langle\psi_{0}^{(2)}| \hat{U}(\tau)|\psi_{3}^{(1)}\rangle |^{2}$. For more details about the proof see \cite{purkait2023PRE}.

Similarly, we can prove for the unitary compression stage that $|\langle\psi_{3}^{(1)}| \hat{V}(\tau)|\psi_{0}^{(2)}\rangle |^{2} =  |\langle\psi_{0}^{(1)}| \hat{V}(\tau)|\psi_{3}^{(2)}\rangle |^{2}$.
Also, using the equivalence between two unitary time evolution operators $\hat{U}(t)$ and $\hat{V}(\tau)$ [see \textbf{App.~\ref{equi}}], we can show that $|\langle\psi_{3}^{(2)}| \hat{U}(\tau)|\psi_{0}^{(1)}\rangle |^{2} =  |\langle\psi_{0}^{(2)}| \hat{U}(\tau)|\psi_{3}^{(1)}\rangle |^{2} = |\langle\psi_{3}^{(1)}| \hat{V}(\tau)|\psi_{0}^{(2)}\rangle |^{2} =  |\langle\psi_{0}^{(1)}| \hat{V}(\tau)|\psi_{3}^{(2)}\rangle |^{2}$.

\section{Derivation of internal energies of a local spin system for quasistatic operation}\label{dev int en  local}


\begin{flushleft}
{\bf At A: }
\end{flushleft}
The density matrix of the local spin at A, $\rho_{AL}=\langle 0_{2}|\rho_{A}| 0_{2}\rangle + \langle 1_{2}|\rho_{A}| 1_{2}\rangle$ is given by 
\begin{equation}
\begin{aligned}
\rho_{AL}&=\frac{1}{2}[p_{0}^{\mathrm{L}} (b_L^2\left|0\right\rangle\left\langle 0\right| + a_L^2 \left|1\right\rangle\left\langle 1\right|) + p_{1}^{\mathrm{L}} (\left|1\right\rangle\left\langle 1\right| + \left|0\right\rangle\left\langle 0\right|) + \\
&~~~~~~p_{2}^{\mathrm{L}} (\left|1\right\rangle\left\langle 1\right| + \left|0\right\rangle\left\langle 0\right|)  + p_{3}^{\mathrm{L}} ( d_L^2\left|0\right\rangle\left\langle 0\right| + c_L^2 \left|1\right\rangle\left\langle 1\right|)]
\;,
\end{aligned}
\end{equation}
where $P_0^L$ and $P_3^L$ are the thermal probabilities of the $0th$ and $3rd$ energy levels at A.

The average internal energy at point A, $\langle E_{A}\rangle_L = \Tr(H_{L1}\rho_{AL})$ is given by
\begin{equation}
\begin{aligned}
&\langle E_{A}\rangle_L = \sum_{j=0,1} \langle j|(-B_L\left|0\right\rangle\left\langle 0\right| + B_L\left|1\right\rangle\left\langle 1\right| )\rho_{LA} |j\rangle\\
& = \frac{B_L}{2}\left[ P_0^L(a_L^2 - b_L^2) + P_3^L(c_L^2 - d_L^2) \right]\\
&= B_L\left[ (P_3^L - P_0^L)(1 - a_L^2) \right]\;,
\end{aligned}
\end{equation}
where we have used $a_L^2 = d_L^2$, $b_L^2 = c_L^2$, and $a_L^2/2 + b_L^2/2 = 1$.

\begin{flushleft}
{\bf At B: }
\end{flushleft}
The density matrix of the local spin at B, $\rho_{BL}=\langle 0_{2}|\rho_{A}| 0_{2}\rangle + \langle 1_{2}|\rho_{A}| 1_{2}\rangle$ is given by 
\begin{equation}
\begin{aligned}
\rho_{BL}&=\frac{1}{2}[p_{0}^{\mathrm{L}} (b_H^2\left|0\right\rangle\left\langle 0\right| + a_H^2 \left|1\right\rangle\left\langle 1\right|) + p_{1}^{\mathrm{L}} (\left|1\right\rangle\left\langle 1\right| + \left|0\right\rangle\left\langle 0\right|) \\
&+ p_{2}^{\mathrm{L}} (\left|1\right\rangle\left\langle 1\right| + \left|0\right\rangle\left\langle 0\right|) + p_{3}^{\mathrm{L}} ( d_H^2\left|0\right\rangle\left\langle 0\right| + c_H^2 \left|1\right\rangle\left\langle 1\right|)]
\;.
\end{aligned}
\end{equation}

The average internal energy at point B, $\langle E_{B}\rangle_L = \Tr(H_{L2}\rho_{LB})$ is give by
\begin{equation}
\begin{aligned}
&\langle E_{B}\rangle_L = \sum_{j=0,1} \langle j|(-B_H\left|0\right\rangle\left\langle 0\right| + B_H\left|1\right\rangle\left\langle 1\right| )\rho_{LB} |j\rangle\\
& = \frac{B_H}{2}\left[ P_0^L(a_H^2 - b_H^2) + P_3^L(c_H^2 - d_H^2) \right]\\
&= B_H\left[ (P_3^L - P_0^L)(1 - a_H^2) \right]\;,
\end{aligned}
\end{equation}
where we have used $a_H^2 = d_H^2$, $b_H^2 = c_H^2$, and $a_H^2/2 + b_H^2/2 = 1$.

\begin{flushleft}
{\bf At C: }
\end{flushleft}
Similarly to point A, we can derive the average internal energy at point C, $\langle E_{C}\rangle_L = \Tr(H_{L2}\rho_{CL})$ given by 
\begin{equation}
\begin{aligned}
&\langle E_{C}\rangle_L  = \frac{B_H}{2}\left[ P_0^H(a_H^2 - b_H^2) + P_3^H(c_H^2 - d_H^2) \right]\\
&= B_H\left[ (P_3^H - P_0^H)(1 - a_H^2) \right]\;,
\end{aligned}
\end{equation}
where $P_0^H$ and $P_3^H$ are the thermal probabilities $0th$ and $3rd$ energy levels at C.

\begin{flushleft}
{\bf At D: }
\end{flushleft}
Similarly to point B, we can derive the average internal energy at D, $\langle E_{D}\rangle_L = \Tr(H_{L1}\rho_{LD})$ given by  
\begin{equation}
\begin{aligned}
&\langle E_{D}\rangle_L = \frac{B_L}{2}\left[ P_0^H(a_L^2 - b_L^2) + P_3^H(c_L^2 - d_L^2) \right]\\
&= B_L\left[ (P_3^H - P_0^H)(1 - a_L^2) \right]\;.
\end{aligned}
\end{equation}

\section{Derivation of internal energies of a local spin system for finite time operation}\label{dev int en local fint}

\begin{flushleft}
{\bf At B: }
\end{flushleft}
The density matrix at B, $\rho_{BL\tau}=\langle 0_{2}|\rho_{A\tau}| 0_{2}\rangle + \langle 1_{2}|\rho_{A\tau}| 1_{2}\rangle$ is given by 
\begin{equation}
\begin{aligned}
\rho_{BL\tau}&=\frac{1}{2}[ p_{1}^{\mathrm{L}} (\left|1\right\rangle\left\langle 1\right| + \left|0\right\rangle\left\langle 0\right|) + p_{2}^{\mathrm{L}} (\left|1\right\rangle\left\langle 1\right| + \left|0\right\rangle\left\langle 0\right|) ] + \\
&P_{0}^L\langle 0_2| \hat{U}(\tau)|\psi_{0}^{(1)}\rangle\langle\psi_{0}^{(1)}|\hat{U}^\dag (\tau)|0_2\rangle + P_{3}^L\langle 0_2| \hat{U}(\tau)|\psi_{3}^{(1)}\rangle \\
&\times \langle\psi_{3}^{(1)}|\hat{U}^\dag (\tau)|0_2\rangle + P_{0}^L\langle 1_2| \hat{U}(\tau)|\psi_{0}^{(1)}\rangle\langle\psi_{0}^{(1)}|\hat{U}^\dag (\tau)|1_2\rangle \\
&~~~~~~~~~~~~~ +~ P_{3}^L\langle 1_2| \hat{U}(\tau)|\psi_{3}^{(1)}\rangle\langle\psi_{3}^{(1)}|\hat{U}^\dag (\tau)|1_2\rangle
\;.
\end{aligned}
\end{equation}

The average internal energy, $\langle E_{LBt}\rangle = \Tr(H_{L2}\rho_{LBt})$ is given by 
\begin{equation}
\begin{aligned}
&\langle E_{B}\rangle_{L\tau} = \sum_{j=0,1} \langle j|(-B_H\left|0\right\rangle\left\langle 0\right| + B_H\left|1\right\rangle\left\langle 1\right| )\rho_{LB} |j\rangle\\
& =  -P^H_0 B_H |\langle00| \hat{U}(\tau)|\psi_{0}^{(1)}\rangle |^{2} -P^H_3 B_H |\langle00| \hat{U}(\tau)|\psi_{3}^{(1)}\rangle |^{2} + \\
&~~~~~~~~~P^H_0 B_H |\langle11| \hat{U}(\tau)|\psi_{0}^{(1)}\rangle |^{2} + P^H_3 B_H |\langle11| \hat{U}(\tau)|\psi_{3}^{(1)}\rangle |^{2}\\
&= B_H (P_3^L - P_0^L) (1 - 2\delta_\tau)\;,
\end{aligned}
\end{equation}
where we have used the microreversibility conditions [for derivation see \textbf{App.~\ref{Micro local spin}}] $|\langle00| \hat{U}(\tau)|\psi_{0}^{(1)}\rangle |^{2} = 1 - \lambda_\tau$, $|\langle00| \hat{U}(\tau)|\psi_{3}^{(1)}\rangle |^{2} = \lambda_\tau$, $|\langle11| \hat{U}(\tau)|\psi_{0}^{(1)}\rangle |^{2} = \lambda_\tau$, $|\langle11| \hat{U}(\tau)|\psi_{3}^{(1)}\rangle |^{2} = 1- \lambda_\tau$.

\begin{flushleft}
{\bf At D: }
\end{flushleft}
Similarly to point B, we can derive the expression of the average internal energy at D, $\langle E_{D}\rangle_{L\tau} = \Tr(H_{L1}\rho_{DL\tau})$ given by  \begin{equation}
\begin{aligned}
&\langle E_{D}\rangle_{L\tau} = B_L (P_3^H - P_0^H) (1 - 2\delta_\tau)\;,
\end{aligned}
\end{equation}
where we need to use the microreversibility conditions [for derivation see \textbf{App.~\ref{Micro local spin}}] $|\langle00| \hat{V}(\tau)|\psi_{0}^{(2)}\rangle |^{2} = 1 - \delta_\tau$, $|\langle00| \hat{V}(\tau)|\psi_{3}^{(2)}\rangle |^{2} = \delta_\tau$, $|\langle11| \hat{V}(\tau)|\psi_{0}^{(2)}\rangle |^{2} = \delta_\tau$, and $|\langle11| \hat{V}(\tau)|\psi_{3}^{(2)}\rangle |^{2} = 1- \delta_\tau$.

\section{Proof of the micro-reversibility condition for the local spin system}\label{Micro local spin}
We can proof the relation $|\langle00| \hat{U}(\tau)|\psi_{3}^{(1)}\rangle |^{2} = |\langle11| \hat{U}(\tau)|\psi_{0}^{(1)}\rangle |^{2}$  using the completeness relation $\sum_{i=0}^3|\psi_i^{(1)}\rangle \langle \psi_i^{(1)}| = \mathbb{I}$ and the conservation of probability $|\langle00| \hat{U}(\tau)|\psi_{0}^{(1)}\rangle |^{2} + |\langle11| \hat{U}(\tau)|\psi_{0}^{(1)}\rangle |^{2} = 1$, whereas other two terms $|\langle01| \hat{U}(\tau)|\psi_{0}^{(1)}\rangle |^{2} = 0$, and $|\langle10| \hat{U}(\tau)|\psi_{0}^{(1)}\rangle |^{2} = 0$.

Similarly, we can prove that $|\langle00| \hat{V}(\tau)|\psi_{3}^{(2)}\rangle |^{2} = |\langle11| \hat{V}(\tau)|\psi_{0}^{(2)}\rangle |^{2}$, where we need to use the conservation of probability $|\langle00| \hat{V}(\tau)|\psi_{0}^{(2)}\rangle |^{2} + |\langle11| \hat{V}(\tau)|\psi_{0}^{(2)}\rangle |^{2} = 1$.


\bibliography{ref}

\end{document}